\documentclass[aps,physrev,twocolumn,superscriptaddress,floatfix]{revtex4-2}

\usepackage{graphicx}

\usepackage{dcolumn}   
\usepackage{bm}        
\usepackage{mathrsfs}  
\usepackage{amsmath}   
\usepackage{ulem}

\usepackage[dvipsnames]{xcolor}
\usepackage{hyperref}
\hypersetup{
    colorlinks = true,
    linkcolor  = Blue,
    urlcolor   = Blue,
    citecolor  = Blue
}
\usepackage{siunitx}
\usepackage{xspace}

\newcommand{\orcid}[1]{\href{https://orcid.org/#1}{\includegraphics[height=\fontcharht\font`\B]{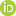}}}

\newcommand{\nsp}{$00~$}
\newcommand{\pp}{$\uparrow\uparrow~$}
\newcommand{\mm}{$\downarrow\downarrow~$}
\newcommand{\nspc}{$00$}
\newcommand{\ppc}{$\uparrow\uparrow$}
\newcommand{\mmc}{$\downarrow\downarrow$}

\def\newacronym#1#2#3{\gdef#1{#3 (#2)\gdef#1{#2}}}

\renewcommand{\newacronym}[3]{%
  \newcommand{#1}[1][]{%
    \renewcommand{#1}[1][]{#2####1\xspace}%
    #3##1 (#2)\xspace%
  }%
}

\newcommand{\etk}{\xspace{\sc EinsteinToolkit}\xspace}

\newcommand{\carpet}{\xspace{\sc Carpet}\xspace}
\newcommand{\cactus}{\xspace{\sc Cactus}\xspace}
\newcommand{\fuka}{\xspace{\sc FUKA}\xspace}
\newcommand{\igm}{\xspace{\sc IGM}\xspace}
\newcommand{\ctg}{\xspace{\sc CTGamma}\xspace}
\newcommand{\thcm}{\xspace{\sc THC\_M1}\xspace}
\newcommand{\thc}{\xspace{\sc THC}\xspace}

\newcommand{\illinois}{\xspace{\sc IllinoisGRMHD}\xspace}

\newcommand{\weakrates}{\xspace{\sc WeakRates}\xspace}
\newcommand{\skynet}{\xspace{\sc SkyNet}\xspace}

\newacronym{\amr}{AMR}{adaptive mesh refinement}
\newacronym{\hrsc}{HRSC}{high resolution shock capturing}
\newacronym{\grmhd}{GRMHD}{general relativistic magnetohydrodynamics}
\newacronym{\srmhd}{SRMHD}{special relativistic magnetohydrodynamics}
\newacronym{\mhd}{MHD}{magnetohydrodynamics}
\newacronym{\imri}{IMRI}{intermediate mass-ratio inspiral}
\newacronym{\imbh}{IMBH}{intermediate mass black hole}
\newacronym{\smbh}{SMBH}{supermassive black hole}
\newacronym{\bbh}{BBH}{binary black hole}
\newacronym{\bhns}{BHNS}{black hole-neutron star}
\newacronym{\bh}{BH}{black hole}
\newacronym{\bns}{BNS}{binary neutron star}
\newacronym{\rns}{RNS}{rotating neutron star}
\newacronym{\hmns}{HMNS}{hypermassive neutron star}
\newacronym{\ns}{NS}{neutron star}
\newacronym{\elmag}{EM}{electromagnetic}
\newacronym{\gw}{GW}{gravitational wave}
\newacronym{\eos}{EOS}{equation of state}
\newacronym{\gr}{GR}{General Relativity}
\newacronym{\grb}{GRB}{gamma-ray burst}
\newacronym{\sgrb}{sGRB}{short gamma-ray burst}
\newacronym{\pn}{PN}{post-Newtonian}
\newacronym{\pde}{PDE}{partial differential equation}
\newacronym{\khi}{KHI}{Kelvin-Helmholtz instability}
\newacronym{\mri}{MRI}{magnetorotational instability}
\newacronym{\kNe}{KN}{kilonova}
\newacronym{\nse}{NSE}{nuclear statistical equilibrium}
\newacronym{\parker}{PI}{Parker instability}

\begin{document}

\title{Magnetic Eruption and Nucleosynthesis in GR\texorpdfstring{$\nu$}{\textit{nu}}MHD Simulations of Spinning Neutron Star Mergers}

\author{Allen Wen\:\orcid{0000-0001-9505-6557}}
\email{acw6923@g.rit.edu}
\affiliation{
Center for Computational Relativity and Gravitation \& School of Physics and Astronomy, Rochester Institute of Technology, 170 Lomb Memorial Drive, Rochester, New York 14623, USA
}

\author{Jay V. Kalinani\:\orcid{0000-0002-2945-1142}}
\affiliation{
The Grainger College of Engineering, Department of Physics \& Illinois Center for Advanced Studies of the Universe, University of Illinois Urbana-Champaign, Urbana, Illinois 61801, USA 
}
\affiliation{
Center for Computational Relativity and Gravitation \& School of Physics and Astronomy, Rochester Institute of Technology, 170 Lomb Memorial Drive, Rochester, New York 14623, USA
}

\author{Michail Chabanov\:\orcid{0000-0001-9676-765X}}
\affiliation{
Center for Computational Relativity and Gravitation \& School of Physics and Astronomy, Rochester Institute of Technology, 170 Lomb Memorial Drive, Rochester, New York 14623, USA
}

\author{Manuela Campanelli\:\orcid{0000-0002-8659-6591}}
\affiliation{
Center for Computational Relativity and Gravitation \& School of Physics and Astronomy, Rochester Institute of Technology, 170 Lomb Memorial Drive, Rochester, New York 14623, USA
}

\author{Riccardo Ciolfi\:\orcid{0000-0003-3140-8933}}
\affiliation{
INAF, Osservatorio Astronomico di Padova, Vicolo dell'Osservatorio 5, I-35122 Padova, Italy }
\affiliation{
INFN, Sezione di Padova, Via Francesco Marzolo 8, I-35131 Padova, Italy
}

\author{Yosef Zlochower\:\orcid{0000-0002-7541-6612}}
\affiliation{
Center for Computational Relativity and Gravitation \& School of Physics and Astronomy, Rochester Institute of Technology, 170 Lomb Memorial Drive, Rochester, New York 14623, USA
}

\begin{abstract}

We present three-dimensional general relativistic magnetohydrodynamics simulations of equal-mass binary neutron star mergers with varied neutron star spin configurations and second-moment neutrino transport, following the formation and early evolution of long-lived remnants. We compare a fiducial irrotational binary with binaries having spins that are aligned or antialigned with the orbital angular momentum, and examime how spin affects the merger dynamics, magnetic field evolution, outflows, and nucleosynthesis. Compared to the fiducial case, the aligned spin configuration releases more cold, neutron-rich tidal ejecta in the equatorial plane, which enables the development of a more tightly collimated polar outflow erupting from the remnant and inner accretion disk. Conversely, the case with spins antialigned with the orbit experiences a more violent collision at merger, disrupting magnetic amplification, loading the environment with debris, and impeding the propagation of magnetically driven winds. Strong neutrino reprocessing of the polar outflow in the irrotational and aligned spin cases produces \texorpdfstring{$2.4\times 10^{-3}\,M_\odot$}{2.4e-3 solar masses} of proton-rich (\texorpdfstring{$Y_e \geq 0.49$}{Ye >= 0.49}) material, resulting in the synthesis of light r-process elements, whose subsequent decay potentially sends a unique electromagnetic signal from long-lived remnants. However, the outflows remain too dense and slow to be consistent with typical short gamma-ray bursts.

\end{abstract} 

\maketitle

\section{Introduction}

The observation of the \bns merger GW170817 firmly established these events as prime multimessenger engines, powering both \gw[s] and a rich array of \elmag transients, including \grb[s] and \kNe[e] \cite{LIGOScientific:2017vwq, LIGOScientific:2017ync, LIGOScientific:2018hze, LIGOScientific:2018cki, LIGOScientific:2017fdd, LIGOScientific:2017zic, cowperthwaite_electromagnetic_2017, ANTARES:2017bia, LIGOScientific:2017pwl, Evans:2017mmy, Margutti:2017cjl, Pian:2017gtc}. However, forging the link between highly nonlinear physics at the merger site and observable \elmag signals remains a formidable challenge. Such a link requires capturing a delicate interplay of general relativity, \mhd, nuclear \eos models, and weak interactions. Notably, magnetic fields are necessary for driving angular momentum transport in the postmerger accretion disk and launching the relativistic jets associated with \grb[s]. Simultaneously, the transfer of energy and lepton number via neutrino radiation cools the remnant and changes the electron fraction ($Y_e$) of the ejecta, which ultimately determines the nucleosynthetic yield and ``color'' of the resulting \kNe \cite{metzger_red_2014, metzger_kilonovae_2017}.

Directly modeling this neutrino radiation requires solving the seven-dimensional Boltzmann transport equations describing the phase space evolution of the neutrino distribution function. This calculation, alongside solving the \grmhd and Einstein equations, imposes high computational demands and is currently impractical without simplifying approximations. One successful approximation is the second-moment formulation, often called an ``M1'' scheme, which evolves the lowest two moments of the distribution function constructed from the neutrino energy density and flux \cite{thorne_relativistic_1981, shibata_truncated_2011, Foucart:2016rxm, radice_new_2022}. Studies coupling M1 neutrino transport to general relativistic hydrodynamics codes are capable of capturing anisotropic features in the radiation field and their dynamic impact on the fluid and its composition, resulting in vastly different accretion disk morphologies, ejecta properties, and \kNe light curves \cite{zappa_binary_2023, espino_impact_2024, bernuzzi_long-lived_2025} compared to less sophisticated leakage or first moment schemes \cite{1996A&A...311..532R, galeazzi_implementation_2013, radice_dynamical_2016, werneck_addition_2023}. Accurate $Y_e$ evolution is critical for predicting the amount of neutron-rich material ejected from \bns mergers, and quantifying their contribution to observed abundances of heavy elements formed from astrophysical rapid neutron capture (r-process) reactions (e.g. \cite{cowan_origin_2021, arnould_r-process_2007}). Additionally, coupled nucleosynthesis and \kNe radiative transfer postprocessing of neutrino irradiated, proton-rich outflows from long-lived remnants in such simulations predicts significant production of $^{56}\mathrm{Ni}$ and $^{56}\mathrm{Co}$, which flattens the decline of the \kNe signal over their nuclear decay timescales of $10 - 100$ days \cite{magistrelli_element_2024, bernuzzi_long-lived_2025, jacobi_56ni_2026}. Thus, fully capturing the neutrino radiation from the remnant provides a new tool to constrain its lifetime and deduce features of the dense matter \eos.

Multiple groups have began to couple M1 schemes to \grmhd codes \cite{schianchi_m1_2023, musolino_practical_2024, daszuta_gr-athena_2026} capable of simulating fully magnetized \bns mergers \cite{sun_jet_2022, musolino_extended_2024, musolino_impact_2025, daszuta_gr-athena_2026, neuweiler_general-relativistic_2026}. Previous \grmhd simulations with no, or less sophisticated, neutrino transport resulting in long-lived remnants have developed strong, large-scale postmerger magnetic field structures extending hundreds of kilometers away from the merger site \cite{ciolfi_general_2019} and collimating a relativistic outflow \cite{ciolfi_collimated_2020, most_impact_2023, aguilera-miret_delayed_2024, bamber_jetlike_2024, jiang_long-term_2025, kalinani_jetenvironment_2026} consistent with high velocities and high $Y_e$ inferred from the optically thin blue \kNe observed with GW170817 \cite{combi_jets_2023, curtis_r-process_2023}. The dynamic interplay between neutrino fields modeled through an M1 scheme and magnetically driven outflows has proven to be essential, as neutrino cooling of the disk will lead to an earlier \parker-driven magnetic eruption \cite{musolino_impact_2025} and alter its composition \cite{neuweiler_general-relativistic_2026, daszuta_gr-athena_2026}.

An additional consideration in this multiphysics environment is the intrinsic spin of the merging \ns[s]. Binary pulsar observations confirm that pre-merger \ns[s] can possess non-negligible spins \cite{lorimer_binary_2008}, and previous \bns merger simulations demonstrate that intrinsic spin profoundly impacts the merger dynamics due to spin-orbit coupling \cite{kastaun_black_2013, bernuzzi_mergers_2014, dietrich_gravitational_2017, kastaun_structure_2017, east_binary_2019, dudi_high-accuracy_2022, ruiz_effects_2019, most_impact_2019, karakas_effect_2026, schianchi_black-hole_2024, ng_initial_2026, neuweiler_general-relativistic_2026}. Binaries with spins aligned with the orbital angular momentum experience a repulsive spin-orbit coupling \cite{kidder_coalescing_1995} leading to a longer inspiral, a stronger tidal interaction, and a cleaner polar environment immediately after merger. Conversely, anti-aligned spins induce a rapid plunge and a violent collision that ejects massive amounts of dynamical ejecta, which can subsequently fall back and resist the propagation of secular outflows. Only one previous study considered \grmhd and M1 simulations of a binary with spins, using an aligned spin case with each \ns having a moderate dimensionless spin of $\chi = J/M^2 = 0.1$ \cite{neuweiler_general-relativistic_2026}.

In this paper, we present 3D \grmhd simulations of highly spinning ($\chi=\pm0.43$) \bns mergers that incorporate M1 neutrino transport, choosing to model \ns[s] with spin periods compatible with the fastest observed pulsar \cite{Hessels2006} to constrain the maximum effects of spin. We systematically investigate the effects of spin on the dynamics of the merger, magnetic eruption, and subsequent $Y_e$ evolution of the ejecta.  We simulate three equal-mass \bns configurations: a nonspinning fiducial case (\nspc), a case with spins aligned with the orbital angular momentum (\ppc), and a case with spins antialigned (\mmc). Utilizing the \illinois (\igm) code \cite{werneck_addition_2023, etienne_illinoisgrmhd_2015} coupled with the \thcm neutrino transport module \cite{radice_new_2022}, we track the postmerger evolution up to $\sim \SI{60}{ms}$ to observe the onset of the \parker and the launching of mildly relativistic, magnetically driven winds. Finally, we provide passive Lagrangian tracer particle trajectories to the \skynet nuclear reaction network \cite{lippuner_skynet_2017} to calculate nucleosynthesis yields, highlighting how the interplay between spin-induced disk dynamics, magnetic outflows, and neutrino irradiation produces distinct nuclear abundances with implications for \kNe signatures, including the enhanced production of light nuclei in the polar outflow.

The paper is organized as follows. In Section \ref{sec:techniques}, we detail our numerical methods, initial data, and grid setup. In Section \ref{sec:results}, we present the dynamics of the merger, gravitational wave emission, magnetic field amplification, and neutrino radiation, followed by a detailed analysis of the ejecta properties and nucleosynthesis yields. We summarize our findings and conclude in Section \ref{sec:summary}.

\section{Techniques}
\label{sec:techniques}

\subsection{Numerical Methods}
To conduct the necessary multiphysics calculations required for \bns merger simulations, we combine several code modules for \mhd, spacetime, and neutrino radiation evolution via the \cactus infrastructure provided by the \etk \cite{maxwell_rizzo_2025_15520463}. The physics modules then perform the simulation on nested \amr grids provided by the \carpet driver \cite{schnetter_evolutions_2004}.

We use the \igm code \cite{etienne_illinoisgrmhd_2015} upgraded with tabulated \eos support \cite{werneck_addition_2023} to update \mhd fields. The public version of \igm is a \hrsc finite volume \grmhd code widely used for astrophysics applications (e.g., \cite{porth_event_2019, ennoggi_relativistic_2025, de_simone_accretion_2025, manikantan_magnetically_2025, ennoggi_merger_2026, raithel_realistic_2021, lopez_armengol_handing_2022, zenati_bound_2023, zenati_dynamics_2024}). \igm employs the PPM method for reconstruction of primitive variables \cite{colella_piecewise_1984}, the Harten-Lax-van Leer approximate Riemann solver \cite{harten_upstream_1983}, and the conservative-to-primitive inversion algorithms of \cite{palenzuela_effects_2015} and \cite{newman_primitive_2014} for tabulated \eos. To maintain the divergence-free nature of the magnetic field, \igm evolves the vector potential using the generalized Lorenz gauge \cite{etienne_relativistic_2012} and computes the magnetic field from its curl.

We use the \thcm code \cite{radice_new_2022} to evolve the energy-integrated zeroth and first moments of the neutrino distribution function while prescribing an analytic closure for the second moment. We consider three independent neutrino species consisting of electron neutrinos ($\nu_e$), electron antineutrinos ($\nu_a$), and a general heavy neutrino field combining muon and tau neutrinos and antineutrinos ($\nu_x$). Notably, the \thcm code includes all nonlinear neutrino-matter interaction terms, improving the realism of trapped neutrino motion in relativistic fluids \cite{foucart_robustness_2024}. These terms become stiff in the optically thick limit, requiring an implicit solve at each substep in the two-step semi-implicit time integration scheme. We couple \mhd and neutrino fields via reaction rates implemented in the \weakrates library included with the \thc codebase (e.g., \cite{galeazzi_implementation_2013, radice_dynamical_2016}) and compute the backreaction of neutrino radiation onto the fluid as:
\begin{equation}
    \nabla_\mu T^{\mu\nu}_{\mathrm{M1}} = -\nabla_\mu T^{\mu\nu}_{\mathrm{MHD}}
    \label{eq:cons_E}
\end{equation}
where $\nabla_\mu$ denotes a covariant derivative, $T^{\mu\nu}_{\mathrm{M1}}$ is the neutrino stress-energy tensor in the M1 formalism, and $T^{\mu\nu}_{\mathrm{MHD}}$ is the matter stress-energy tensor. \thcm also evolves the number density of each neutrino species to model the impact of neutrino emission and absorption on the local electron fraction $Y_e$ via conservation of lepton number. To ensure stable evolution of the M1 state vector, the \thcm algorithm hybridizes a high-order flux with a diffusive Lax-Friedrichs flux in numerically problematic regions of the solution, including where a high-frequency odd-even decoupling is detected (see discussion of Eq.~(42) in \cite{radice_new_2022}). We find that introducing additional Kreiss-Oliger dissipation in the presence of this decoupling stabilizes radiation advection in optically thin regions and removes numerical artifacts when radiation crosses refinement boundaries in vertex-centered \amr grids. This is the only modification we make to the core implementation of \thcm. We note that the use of an energy-integrated, or gray, scheme cannot fully capture the true neutrino distribution function in semitransparent regimes, leading to significant differences in reaction rates \cite{alford_thermalization_2026} and neutrino-matter interactions in the disk and outflow \cite{cheong_energy-dependent_2024} when compared to methods modeling a nontrivial neutrino energy distribution.

For spacetime evolution, we use the Z4c \cite{bernuzzi_constraint_2010, hilditch_compact_2013} formulation of the Einstein field equations implemented in the \ctg code \cite{pollney_high_2011, reisswig_formation_2013}.

Spacetime and \grmhd fields are evolved using a fourth-order Runge-Kutta (RK4) scheme, while M1 fields are evolved in a separate second-order semi-implicit step. We use a Courant-Friedrichs-Lewy (CFL) factor of $0.35$.

Additional details about methods used for analysis and postprocessing are found in Appendix \ref{ax:analysis}.

\subsection{Initial Data}
\label{ssec:id}
The initial neutron stars are modeled with the APRLDP \eos \cite{akmal_equation_1998, schneider_akmal-pandharipande-ravenhall_2019}, which provides a robust description of dense nuclear matter consistent with current astrophysical constraints. We choose to initialize equal mass binaries in three spin configurations: a fiducial nonspinning case, a case with spins aligned with the orbital angular momentum such that \ns dimensionless spins are $\chi_\mathrm{NS1} = \chi_\mathrm{NS2} = +0.43$, and a case with antialigned spins such that $\chi_\mathrm{NS1} = \chi_\mathrm{NS2} = -0.43$. We will refer to these cases as \nspc, \ppc, and \mmc, respectively. Spin periods are calibrated to match the largest observed pulsar frequency of $716$ Hz for PSR J1748–2446ad \cite{Hessels2006}. Each \ns has a gravitational mass of $M = 1.35 M_\odot$. We use the \fuka solver \cite{papenfort_new_2021} to generate initial data from a constant temperature  ($T=\SI{0.01}{MeV}$), $\beta$-equilibrated slice of the 3D APRLDP \eos table. As stated in the introduction, spin-orbit coupling changes the rate at which the binary loses angular momentum, thus changing the inspiral duration when starting from a fixed separation. To induce comparable inspiral durations across the three cases, we initialize the \nsp and \pp binaries at a coordinate separation of $\SI{45}{km}$, and the \mm binary at a coordinate separation of $\SI{55}{km}$. 

For the \nsp and \pp simulations, we superimpose dipolar magnetic fields within the \ns interiors at $t=0$, using a toroidal vector potential
\begin{equation}
    A_\phi = A_b R^2 \max(P - P_\mathrm{cut}, 0)^2
    \label{eq:init_A}
\end{equation}
where $A_b$ is a parameter determining the overall strength of the magnetic field, $R = \sqrt{(x-x_\mathrm{CoM})^2 + (y - y_\mathrm{CoM})^2}$ is the cylindrical radius relative to a star's center of mass, $P$ is the fluid pressure, and $P_\mathrm{cut}$ is a parameter set to $0.4\max(P)$. We choose $A_b$ such that the total initial magnetic energy
\begin{equation}
    E_\mathrm{mag} = \frac{1}{2}\int u^t\,b^2\,\sqrt{-g}\,d^3x
    \label{eq:Emag}
\end{equation}
is $E_\mathrm{mag} = \SI{2.5e49}{erg}$ and $B_\mathrm{max} = \SI{3.6e16}{G}$, where $u^t$ is the time component of the fluid 4-velocity, $g$ is the metric determinant, $b$ is the magnitude of the comoving magnetic field, and $B_\mathrm{max}$ refers to the maximum magnetic field strength on the grid. While these values are much higher than magnetic fields of $10^7 - 10^{11}$ G inferred from binary pulsar observations \cite{lorimer_binary_2008}, they are necessary to overcome the effects of finite grid resolution limiting the efficiency of turbulent magnetic amplification at merger, and serve to approximate realistic postmerger field topologies from high resolution \bns simulations \cite{Palenzuela2022, kiuchi_large-scale_2024, gutierrez_turbulent_2026}. This approximation has known drawbacks. Low resolution grids are unable to fully break down the initial field's dipolar structure into a small scale, turbulent state consistent with high resolution simulations, and subsequently underestimate the time required to reorganize the turbulent field into larger structures capable of affecting postmerger dynamics \cite{aguilera-miret_delayed_2024}. Our use of $z$-reflection symmetry (see Section \ref{ssec:grid}) may exacerbate this issue by enforcing structure in the field along the equatorial plane \cite{gutierrez_magnetic_2025}.

The \mm case required a modified prescription due to more complex dynamics present in the inspiral. The spin magnitude is high enough to shift $f$-mode frequencies of the \ns[s] to lower values of $f_f \approx \SI{1370}{Hz}$ comparable to the late inspiral \gw frequency, such that resonant excitation occurs \cite{kuan_tidal_2025}. Note that \cite{kuan_tidal_2025} isolated the full linear growth of this resonance by simulating a binary with a lower \ns $f_f$ over $\sim 15$ orbits from a large initial separation. While the resonance in our case must occur at closer separations where additional nonlinear merger dynamics become important, we nevertheless observe this process to drive higher order modes to deform the stars and trigger internal motions during the inspiral, rearranging and amplifying the initial magnetic field configuration. While this tidal resonance is physical, the initial magnetic field from Eq.~(\ref{eq:init_A}) is ad hoc, and thus the resulting field generated from these tidal interactions is not necessarily informative. Thus, to better facilitate a comparison to the \nsp and \pp cases where the initial magnetic field remains relatively constant during the inspiral, we instead superimpose the magnetic field from Eq.~(\ref{eq:init_A}) $\SI{0.8}{ms}$ before merger in the \mm case, using $P_\mathrm{cut} = 10^{-8} \max(P)$ and setting $A_b$ such that the total $E_\mathrm{mag}$ at merger is the same across all cases (see Fig.~\ref{fig:khi}). We also reset the neutrino fields to $\beta$-equilibrium \SI{0.2}{ms} before merger to maintain stability in the M1 algorithm throughout the more violent merger and extreme thermodynamics that result from the preceding rapid plunge. Additional information about our estimate of $f_f$ and the conditions prior to merger in the \mm case can be found in Appendix \ref{ax:mm}.

\subsection{Grid Setup}
\label{ssec:grid}
We construct a large \carpet grid with an extent of $[-51500,  51500] \times [-51500, 51500] \times [0, 51500]$ km in the x, y, and z directions, enforcing reflection symmetry across the x-y plane. This large domain allows us to capture the entire outflow in preparation for a follow-up study where we hand off data to a \srmhd code to follow unbound material for longer time and length scales as in \cite{pavan_jet-environment_2023, pavan_role_2025, dreas_simulating_2026}. We bridge the scale separation between the dynamical ejecta front and the small scale turbulence at the merger site with 12 levels of mesh refinement. The grid spacing decreases by a factor of two with each successive refinement level, culminating with a standard resolution of $\Delta x_\mathrm{SR} = \SI{177}{m}$ in the finest level. To evaluate resolution effects, we also simulate the \nsp case at a low resolution of $\Delta x_\mathrm{LR} = \SI{223}{m}$ and compare the two resolutions in Appendix \ref{ax:res}. Two refinement centers track the \ns[s] during the inspiral phase and are replaced by a single refinement center at the origin once the \ns[s] reach a separation of $\SI{13.3}{km}$.

\subsubsection{Low-Density Atmosphere}
\label{sssec:atm}
Ideal \grmhd codes often require a numerical density floor, or atmosphere, as the \hrsc schemes employed cannot evolve a pure vacuum. To ensure free propagation of dynamical ejecta at the large distances we aim to simulate, we prescribe a radially decreasing density floor
\begin{equation}
    \rho_\mathrm{atm}(r) = 
    \begin{cases}
        \rho^* & r \leq r^* \\
        \max\left(\rho^* \left(\frac{r^*}{r}\right)^{n_\rho},\,\min(\rho_\mathrm{EOS})\right) & r > r^*
    \end{cases}
    \label{eq:graded_atm}
\end{equation}
where $\rho^*,\, n_\rho,$ and $r^*$ are free parameters that set the atmosphere density $\rho^*$ in a central region of radius $r^*$, and the power law decay rate $n_\rho$. Following the atmosphere prescription in \cite{kalinani_jetenvironment_2026}, we set $\rho^* = \SI{6.17e4}{g~cm}^{-3},\, n_\rho = 6,$ and $r^* = \SI{74}{km}$.

As noted in Eq.~(\ref{eq:graded_atm}), the minimum density available to the simulation is still limited by the minimum density in the \eos table $\min(\rho_\mathrm{EOS})$. To construct an \eos table covering the necessary density range, we first use the {\sc SROEOS} code provided by \cite{schneider_open-source_2017, schneider_akmal-pandharipande-ravenhall_2019} to generate a table following the APRLDP model covering $\rho \in [113,\, 1.05\times10^{16}]~\mathrm{g~cm}^{-3},\, T \in [7.3 \times 10^{-4},\, 251]~\mathrm{MeV},$ and $Y_e \in [0.005, 0.655]$. We then use the method described in \cite{hayashi_general-relativistic_2022} to append values from the Timmes \eos \cite{timmes_accuracy_2000} to the low density end of the table, extending $\min(\rho_\mathrm{EOS})$ to $10^{-3}~\mathrm{g~cm}^{-3}$. All simulations then close the \grmhd equations according to this extended \eos table.

Both the APRLDP and Timmes \eos models include a trapped photon component with specific internal energy
\begin{equation}
    \epsilon_\mathrm{rad}=\frac{4\sigma_\mathrm{SB}}{c} \frac{T^4}{\rho}
    \label{eq:eps_rad}
\end{equation}
where $\sigma_\mathrm{SB}$ is the Stefan-Boltzmann constant and $c$ is the speed of light. At fixed $T$, $\epsilon_\mathrm{rad}$ increases at low $\rho$, which leads to numerical issues if this large value is enforced as the atmospheric $\epsilon$. To mitigate this issue, we also prescribe a radially decreasing $T_\mathrm{atm}$
\begin{equation}
    T_\mathrm{atm}(r) = 
    \begin{cases}
        T^* & r \leq r^* \\
        \max\left(T^* \left(\frac{r^*}{r}\right)^{n_T},\,\min(T_\mathrm{EOS})\right) & r > r^*
    \end{cases}
    \label{eq:graded_Tatm}
\end{equation}
We find that our choice of $\min(T_\mathrm{EOS})$ stated above, $T^* = \SI{0.01}{MeV},$ and $n_T = 0.4$ are sufficient for ensuring stable evolution of low density material to large distances.
  
\begin{figure*}
    \centering
    \includegraphics[width=\textwidth]{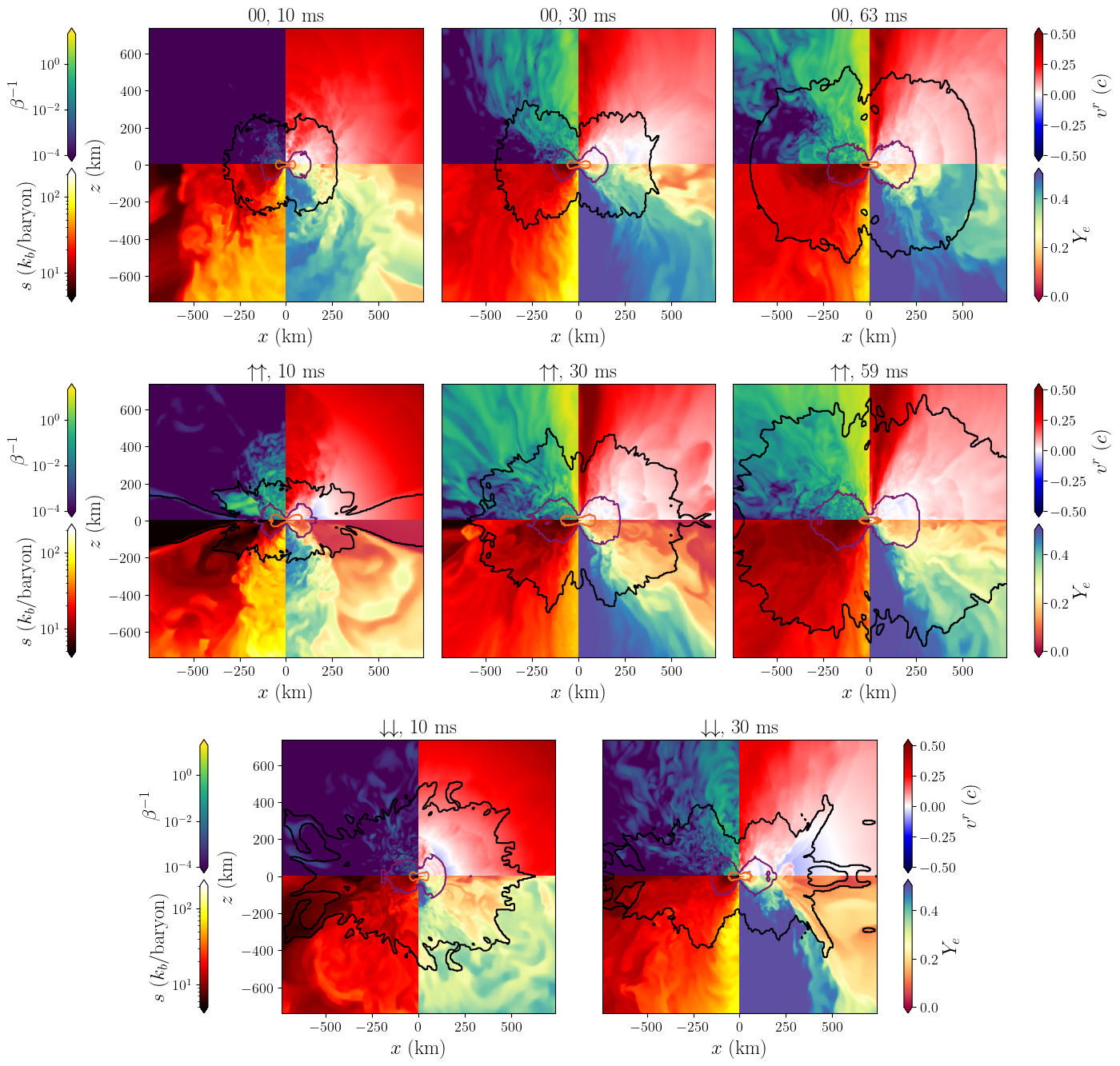}
    \caption{Snapshots in the x-z plane of, clockwise from upper left, the inverse plasma $\beta$ parameter, radial velocity $v^r$, electron fraction $Y_e$, and entropy per baryon $s$ at $\SI{10}{ms},\,\SI{30}{ms},$ and/or the final snapshot of each spin case. The 00, \ppc, and \mm cases are shown in the top, middle, and bottom rows. Contours mark densities of $10^7, 10^9,$ and $10^{11}\,\mathrm{g~cm}^{-3}$. The initial postmerger conditions induced by different merger dynamics are visible in the first snapshots at $\SI{10}{ms}$ and exert strong influences on later dynamics shown to the right.}
    \label{fig:allxz}
\end{figure*}

\section{Results}
\label{sec:results}

\subsection{Evolution Overview}
\label{sec:overview}

The inspiral in each simulation lasts for $\sim5-6$ orbits, or $17.5 - \SI{20.4}{ms}$. While we do not explicitly control the initial separation to measure consequences of spin-orbit coupling, our results are qualitatively consistent with previous post-Newtonian calculations \cite{kidder_coalescing_1995} and numerical relativity simulations of spinning binaries \cite{campanelli_spinning-black-hole_2006, kastaun_black_2013, bernuzzi_mergers_2014, dietrich_gravitational_2017, kastaun_structure_2017, east_binary_2019, dudi_high-accuracy_2022, karakas_effect_2026}: starting from the same initial separation of $\SI{45}{km}$, the \pp inspiral takes $\SI{2.8}{ms}$ longer than the \nsp case, and despite starting from a larger initial separation of $\SI{55}{km}$, the \mm inspiral only lasts $\SI{1.5}{ms}$ longer than the \nsp case. This indicates a slower (faster) loss of angular momentum in the \pp (\mmc) case compared to the \nsp case.

 The \nspc, \ppc, and \mm simulations evolve to 63, 59, and 30 ms postmerger, respectively, with the merger time set to coincide with the maximum \gw strain. Representative snapshots of each spin case are shown in Fig.~\ref{fig:allxz}, displaying how differences in orbital dynamics and the initial collision at merger propagate to the postmerger environment. Each panel in Fig.~\ref{fig:allxz} shows, clockwise from upper left, the inverse plasma parameter $\beta^{-1} = b^2/2P$, the radial velocity $v^r$, the electron fraction $Y_e$, and the entropy per baryon $s$. Qualitatively, the \nsp case, shown in the top row of Fig.~\ref{fig:allxz}, exhibits phenomenology similar to previous ab-initio \grmhd simulations of long-lived \bns merger remnants \cite{ciolfi_general_2017, ciolfi_general_2019, ciolfi_collimated_2020, combi_jets_2023, kiuchi_large-scale_2024, neuweiler_general-relativistic_2026}: magnetic fields amplify within the remnant and accretion disk until magnetic buoyancy overcomes fluid pressure, lifting and collimating a magnetic tower along the polar axis in a process accelerated by neutrino cooling of the disk and neutrino-driven winds clearing polar debris \cite{sun_jet_2022, musolino_impact_2025, jiang_long-term_2025}. The ratio of magnetic to fluid pressure given by $\beta^{-1}$ ranges from $\sim 10^{-4}$ in the dense disk to $\sim 1-10$ in the polar magnetic wind, indicating the dynamic importance of magnetic fields. This accelerates a polar outflow to mildly relativistic speeds of $v^r \sim 0.4-0.5$. Neutrino irradiation from the remnant onto the outflow skews the composition to high $Y_e$ values \cite{combi_jets_2023, curtis_r-process_2023}, powering a ``blue'' \kNe component as opposed to a ``red'' one expected from neutron-rich ejecta \cite{cowperthwaite_electromagnetic_2017}. 

\begin{figure}
\centering
\includegraphics[width=0.9\linewidth]{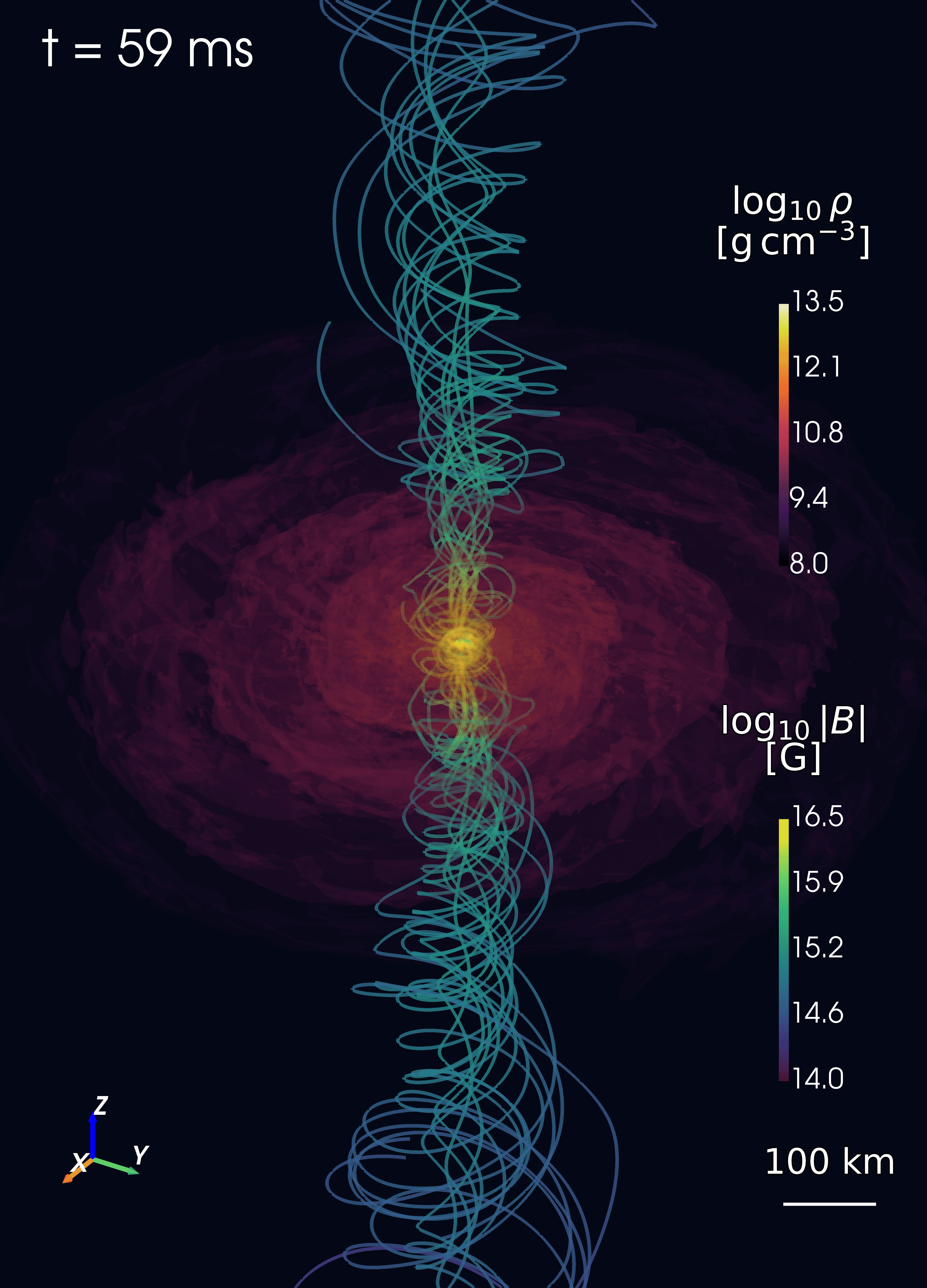}
\caption{Magnetic field line structure for the \pp case at $t=\SI{59}{ms}$ postmerger, with characteristic strengths of order $10^{15}$ G. Density contours show the remnant and equatorial disk at the base of a collimated magnetic tower.}
\label{fig:jet_pp_3D}
\end{figure}

In contrast, the orbital dynamics in the spinning cases sets the stage for unique postmerger dynamics by varying the ``violence" of the collision, a qualitative measure of how quickly the \ns[s] plunge in the final moments of the inspiral, and how strong the resulting hydrodynamic impact, shock heating, and disruption of stellar material is. This violence is reflected in the extent of the dynamical ejecta debris clouds in the leftmost plots of Fig.~\ref{fig:allxz}, the amplitude of density and temperature oscillations following the merger (see Fig.~\ref{fig:khi} and Section \ref{sssec:khi}), and the resulting temperature within the remnant (see Fig.~\ref{fig:T_Rnu} and Section \ref{ssec:neutrino}). The slower inspiral and elongated \ns shape in the \pp case strengthens their tidal interaction prior to merger \cite{dietrich_gravitational_2017, kastaun_structure_2017, east_binary_2019, dudi_high-accuracy_2022, karakas_effect_2026}. The resulting disruption of the \pp \ns[s] and less violent collision extends tidal arms into the equatorial plane, leaving less material polluting the polar axis at early times, shown by density contours in the left column, middle row of Fig.~\ref{fig:allxz}, and a cooler disk throughout the simulation, shown by the $s$ panels. The reduced pollution in the polar region  encourages the breakout and propagation of a fast polar outflow with distinct thermodynamic conditions from the accretion disk \cite{ruiz_effects_2019, bamber_jetlike_2024}. Much like the \nsp case, the polar outflows are driven by the helical magnetic tower threading the polar funnel illustrated in Fig.~\ref{fig:jet_pp_3D}, with characteristic field strengths of order $10^{15}\,\mathrm{G}$. In contrast, this tower forms much earlier (see Section \ref{sssec:eruption}), energizing the outflow to attain higher values of $\beta^{-1}$ and $v_r$, with a stark contrast in $Y_e$ and $s$ between the disk and z-axis. The tidal arms redistribute magnetic energy and angular momentum from the inner remnant to the disk, resulting in less effective magnetic winding within the remnant but greater resolved \mri activity in the disk, inflating the disk at later times. In the \mm case, the \ns[s] experience a more rapid plunge and violent collision at merger, launching a fast, isotropic cloud of dynamical ejecta \cite{most_impact_2019, dudi_high-accuracy_2022} mixed with low angular momentum bound debris, visible through density contours in the left column, bottom row of Fig.~\ref{fig:allxz}. The bound debris falls back towards the remnant, indicated by regions of negative $v_r$. 

\subsection{Gravitational Waves}

\begin{figure}
    \centering
    \includegraphics[width=\linewidth]{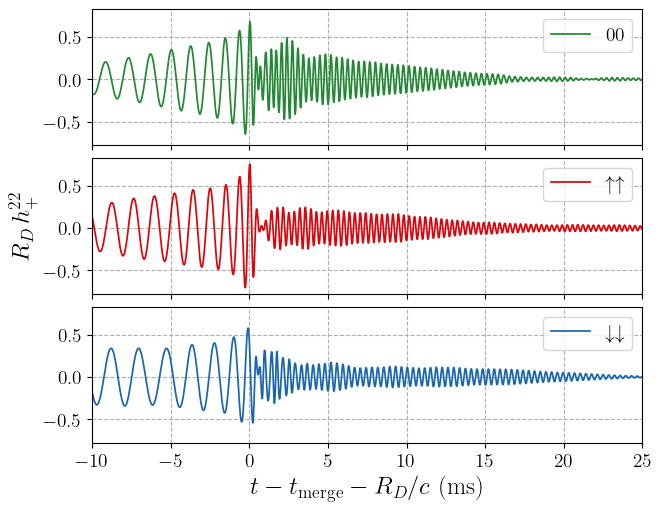}
    \caption{The $h_+$ polarization of the $(2,2)$ \gw mode extracted from a spherical detector, scaled by the detector radius of $R_D = \SI{740}{km}$. The upper, middle and lower panels show the \nspc, \ppc, and \mm cases. The time axis is first shifted by the transmission time from the origin to the detector, and again such that the merger time coincides with the peak strain amplitude.}
    \label{fig:gws}
\end{figure}

\begin{figure}
    \centering
    \includegraphics[width=\linewidth]{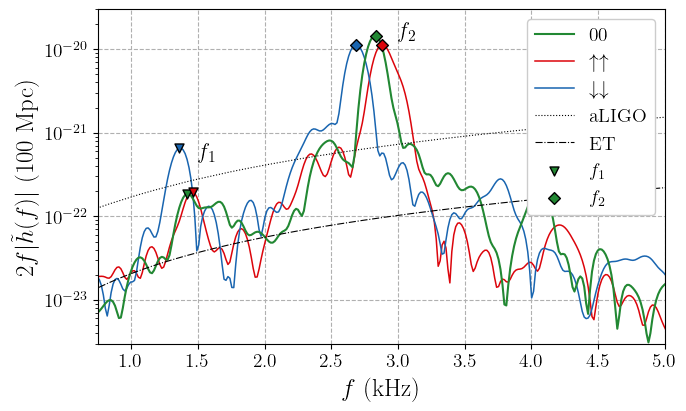}
    \caption{Characteristic frequency strain measured at $\SI{100}{Mpc}$ of the combined $l=2,3$ postmerger \gw strain for each spin case. Triangle and diamond markers indicate characteristic first and second harmonics of the postmerger \gw spectrum. We superimpose sensitivity curves of Advanced LIGO \cite{Evans_Sturani_Vitale_Hall_2020} and the proposed Einstein Telescope \cite{hild_sensitivity_2011}.}
    \label{fig:gw_ffts} 
\end{figure}

We show the \gw signal measured from the dominant $(l=2,\,m=2)$ mode for each spin case in Fig.~\ref{fig:gws}. The merger time has been shifted to $t=0$ such that times $<0$ correspond to the inspiral signal. Qualitative features in the waveforms reflect the overview above. As the \pp (\mmc) case inspirals slower (faster) than the \nsp case, it reaches a higher (lower) orbital frequency at a closer (larger) separation before the nonlinear plunge at merger. The \nsp case exhibits the highest \gw luminosity immediately postmerger, as the tidal disruption in the \pp case and direct collision in the \mm case serve to disrupt the nonaxisymmetric quadrupolar density mode (``bar-mode'') largely responsible for postmerger \gw emission. 

We observe consequences of initial spins that change the \gw frequency spectrum. In Fig.~\ref{fig:gw_ffts}, we show the characteristic frequency strain \cite{moore_gravitational-wave_2014} of all $l=2,3$ \gw modes taken from $t \in [0.5,\,28]$ ms postmerger. We mark peaks associated with the characteristic postmerger frequencies $f_1$ and $f_2$ with triangles and diamonds, respectively. These frequencies are higher (lower) in the \pp (\mmc) case compared to the \nsp case, as angular velocity profiles show the central bulk of the remnant to spin faster (slower) (see Fig.~\ref{fig:Omg_Bphi} and Section \ref{sssec:mag_winding}). We observe a small shift in the $f_1$ frequency which potentially describes an $m=1$ asymmetry driving spiral-wave mass ejection \cite{radice_one-armed_2016}, with $f_1$ ranging from $\SI{1.359}{kHz}$ in the \mm case to $\SI{1.461}{kHz}$ in the \pp case. Consistent with higher amplitude strain in the postmerger $(l=2,\,m=2)$ \gw signal, the dominant $f_2$ peak at $\sim 2f_1$ is highest in the \nsp case. This frequency ranges from $\SI{2.688}{kHz}$ in the \mm case to $\SI{2.878}{kHz}$ in the \pp case. We note that the magnitude of this shift is comparable to the effect of additional microphysical effects in the \eos, e.g. the inclusion of weak-interaction driven viscous effects \cite{chabanov_impact_2025} captured with an M1 scheme \cite{espino_neutrino_2024}, or the presence of a quark-hadron phase transition \cite{Prakash2021}. Thus, high spins can provide an alternative, or complementary, interpretation to future observations of postmerger frequency spectra. As discussed in Appendix \ref{ax:res}, we caution that the \nsp configuration at LR does not exhibit a clear $f_1$ peak and shifts the $f_2$ peak $\SI{35}{Hz}$ higher, hinting that the physics behind the $f_1$ peak is not fully resolved, and higher resolution simulations are required to isolate spin and resolution effects in the postmerger \gw spectrum.

\subsection{Magnetic Amplification}
\label{ssec:mag}

Previous \bns merger simulations featuring long-lived remnants have demonstrated multiple channels responsible for amplifying weak initial magnetic fields to strong, large-scale postmerger structures capable of extending hundreds of kilometers from the merger site \cite{ciolfi_general_2019}, launching relativistic magnetar outflows \cite{ciolfi_collimated_2020, combi_jets_2023, curtis_r-process_2023, most_impact_2023, aguilera-miret_delayed_2024, bamber_jetlike_2024, kalinani_jetenvironment_2026}, and triggering dynamic feedback onto the \hmns via magnetic braking \cite{baumgarte_maximum_1999, shapiro_differential_2000}. Across different spin cases, we observe varying degrees of activity for several of these processes. As the surfaces of the two \ns[s] make initial contact, the shear layer that forms between them is susceptible to the \khi, which triggers a turbulent cascade to viscous scales capable of amplifying magnetic field strengths by several orders of magnitude \cite{kiuchi_efficient_2015, kiuchi_large-scale_2024, gutierrez_turbulent_2026}. After the \khi seeds the remnant with strong, but small scale and incoherent field lines, differential rotation within the remnant winds the field into large scale toroidal structures. In the disk, poloidal magnetic field triggers the \mri \cite{balbus_powerful_1991}, leading to turbulent transport of magnetic fields and angular momentum, and potentially supporting continuous magnetic amplification via an $\alpha\Omega$ dynamo \cite{kiuchi_large-scale_2024}. In concert, these processes enable the resulting magnetic field configuration to break out of the remnant and shape observable signals from the postmerger system.

\subsubsection{Kelvin-Helmholtz Instability}
\label{sssec:khi}

\begin{figure}
    \centering
    \includegraphics[width=0.85\linewidth]{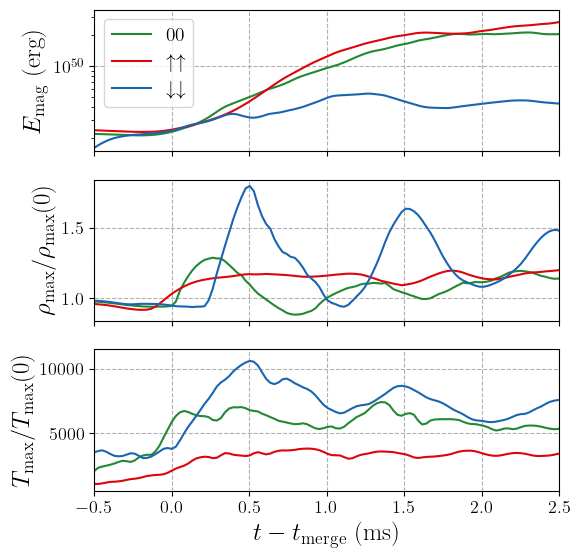}
    \caption{Evolution of total magnetic energy given by Eq.~(\ref{eq:Emag}) (top) and maximum density (middle) and temperature (bottom) normalized by their initial values, shown for the first $\SI{2.5}{ms}$ postmerger.}
    \label{fig:khi}
\end{figure}

\begin{figure}
    \centering
    \includegraphics[width=\linewidth]{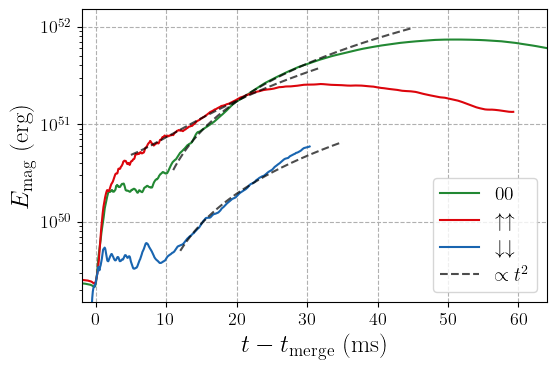}
    \caption{Evolution of the \elmag energy integrated over the simulation domain for each spin case. Black dashed lines represent the contribution of the linear growth of $B_\phi$ due to magnetic winding to the total $E_\mathrm{mag}$.}
    \label{fig:winding}
\end{figure}

\begin{figure*}
    \centering
    \includegraphics[width=\textwidth]{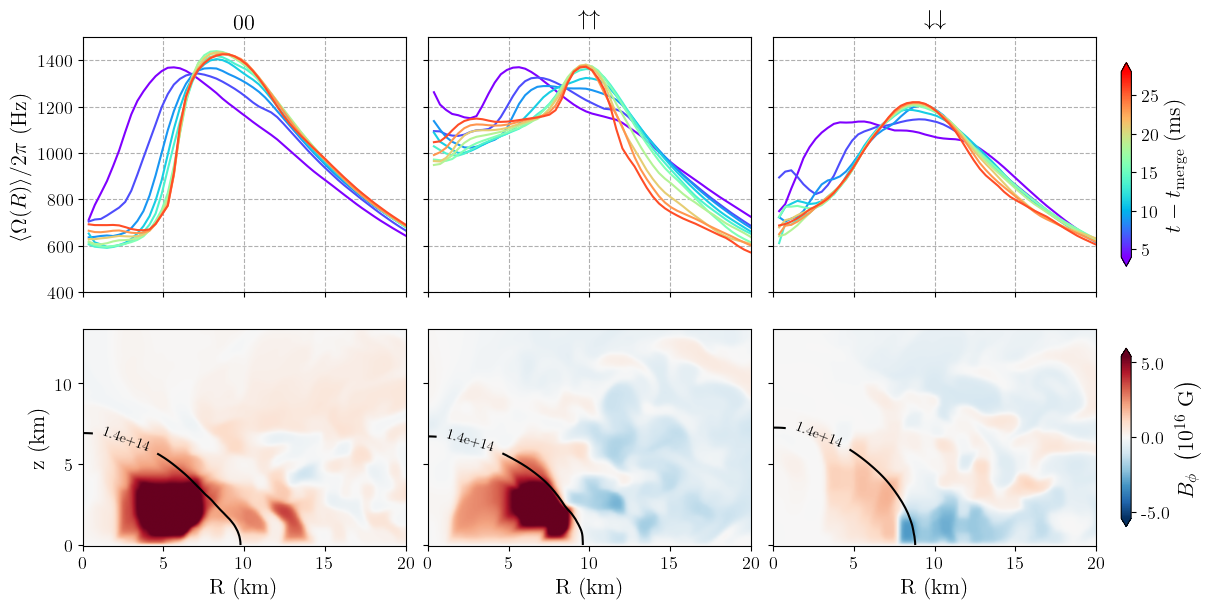}
    \caption{Azimuthally-averaged angular velocity $\Omega$ profiles calculated via Eq.~(\ref{eq:phiavg}) from $4 - \SI{28}{ms}$ postmerger (top) and the toroidal component of the magnetic field $B_\phi$ (bottom) across remnant radii. We show the \nspc, \ppc, and \mm cases from left to right. We take snapshots of $B_\phi$ at $\SI{28}{ms}$ postmerger, coinciding with the red profiles in the top row. A density contour at $\SI{1.4e14}{g~cm}^{-3}$ approximates the remnant surface.}
    \label{fig:Omg_Bphi}
\end{figure*}

Magnetic amplification immediately begins via \khi turbulence. In the \nsp and \pp cases, this efficiently increases the total magnetic energy by an order of magnitude until saturating near $\SI{2e50}{erg}$, shown in the top panel of Fig.~\ref{fig:khi}. This turbulent phase lasts for $\sim \SI{2}{ms}$ before the shear layer dissipates and reorganizes into a hot envelope surrounding the central remnant, a colder region formed from the gradual merging of the initial \ns cores \cite{Hanauske2017}. However, in the \mm case, the two \ns cores directly collide into each other, destroying the shear layer between them. The rapid core collision coincides with the sharp peak in maximum density and temperature in the lower panels of Fig.~\ref{fig:khi} and the flattening of $E_\mathrm{mag}$ in the upper panel at $\sim \SI{0.5}{ms}$ postmerger shown in the blue curves. This is also reflected in the qualitatively different shapes of azimuthally-averaged temperature profiles in Fig.~\ref{fig:T_Rnu}. They show the cold cores in the \nsp and \pp cases with $T \sim 5-\SI{10}{MeV}$, increasing to a maximum at a radius of $\sim \SI{8}{km}$. In contrast, widespread violent shock heating in the \mm case increased temperatures throughout the remnant, with the maximum temperature of $\SI{60}{MeV}$ located at the origin. We reserve a more detailed discussion on $\langle T(R,t)\rangle$ profiles and their link to neutrino emission for Section \ref{ssec:neutrino}.

We note that extreme resolutions of $\Delta x < \SI{10}{m}$ are required to fully resolve the \khi \cite{kiuchi_large-scale_2024, gutierrez_turbulent_2026}, and no study to date has evolved spinning \bns mergers at such a resolution, leaving the true impact of the \khi uncertain. A higher resolution study ($\Delta x = \SI{35}{m}$) of similar initial spin configurations in \cite{ng_initial_2026} observed the opposite hierarchy of \khi efficiency shown in Fig.~\ref{fig:khi}, reporting that stronger shocks and velocity gradients across the \ns contact interface in the \mm case resulted in greater magnetic field amplification relative to \nsp and \pp configurations. However, our results are consistent with \cite{neuweiler_general-relativistic_2026} which used a resolution of $\Delta x = \SI{93}{m}$ to simulate \nsp and \pp cases with lower spins, and reported slightly higher $E_\mathrm{mag}$ following the \khi phase in the \pp case.

\subsubsection{Remnant Rotation and Magnetic Winding}
\label{sssec:mag_winding}

\begin{figure}
    \centering
    \includegraphics[width=0.9\linewidth]{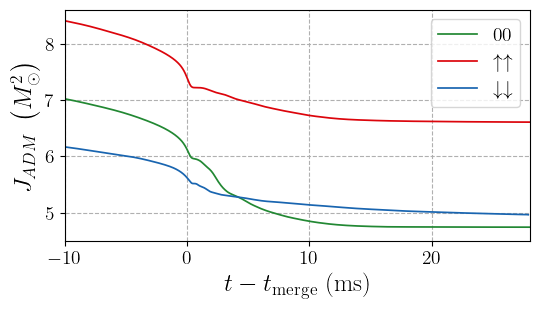}
    \caption{ADM angular momentum $J_{ADM}$ in the late inspiral and early postmerger in all spin cases. We calculate $J_{ADM}$ by subtracting the angular momentum radiated away by \gw[s] from its initial value. Angular momentum loss is measured at $\SI{740}{km}$.}
    \label{fig:JADM}
\end{figure}

Following the initial turbulent phase, the \hmns remnants relax towards an axisymmetric state with a radially varying angular velocity profile \cite{cassing_realistic_2024}. This differential rotation increases the strength of the toroidal magnetic field linearly in time, corresponding to $E_\mathrm{mag} \propto t^2$ shown by black dashed curves in Fig.~\ref{fig:winding}. We observe initial spins to have a strong effect on the rate and saturation of the winding process. First, the rate is determined by the degree of differential rotation within the remnant, shown in the averaged angular velocity $\langle \Omega(R,\,t)\rangle$ from the top row of Fig.~\ref{fig:Omg_Bphi}, computed using Eq.~(\ref{eq:phiavg}). The redistribution of hot, shocked material from the \khi-unstable shear layer between the merging \ns[s] into a rapidly rotating envelope around a slowly rotating core endows the \nsp remnant with the highest angular velocity gradient across its radius, and the fastest magnetic winding. Remnant rotation in the spinning cases is consistent with the evolution of ADM angular momentum $J_{ADM}$ estimated in Fig.~\ref{fig:JADM} by $J_{ADM}(t) = J_{ADM,\,0} - J_{GW}$, where the initial value $J_{ADM,\,0}$ is calculated by the initial data solver \cite{papenfort_new_2021} and $J_{GW}$ is the $z$-component of the radiated angular momentum. In the \pp case, the original \ns cores are initially corotating with the orbit, and thus also with the resulting flow of the remnant and accretion disk. This leads to a more rapidly rotating core, smaller gradient in $\langle \Omega(R, t)\rangle$, and slower magnetic winding, consistent with attaining the highest values of $J_{ADM}$ across all cases. While the \mm \bns has the lowest initial $J_{ADM}$, the \nsp case radiates away more angular momentum through \gw[s], reaching a lower $J_{ADM}$ than the \mm case within $\SI{5}{ms}$ postmerger. We accordingly observe a higher core rotation frequency in the \mm remnant compared to the \nsp case, leading to an intermediate $\langle \Omega(R, t)\rangle$ gradient and winding rate compared to the other cases.

Strong magnetic fields can backreact against the fluid via magnetic braking, slowing rotation and eventually terminating magnetic winding inside the remnant \cite{baumgarte_maximum_1999, shapiro_differential_2000, duez_evolution_2006}. Figure~\ref{fig:winding} shows this in the \nsp and \pp cases through a deviation of $E_\mathrm{mag}(t)$ from the $\propto t^2$ curve followed by a decline when other magnetic field growth channels fail to outpace losses due to conversion to kinetic energy in the outflow and numerical dissipation. The bottom row of Fig.~\ref{fig:Omg_Bphi} shows the toroidal field strength and polarity in an azimuthal slice of the remnant at $t=\SI{28}{ms}$. Note that regions of strong toroidal winding coincide with the location of the steepest gradient in the $\langle \Omega(R,t=\SI{28}{ms})\rangle$ profile, which, in the \pp case, is closer to the surface of the remnant, and thus at a lower density by a factor of $\sim 2$. Thus, as magnetic braking acts on Alfv\'{e}n timescales \cite{baumgarte_maximum_1999}:
\begin{equation}
    \tau_b \sim \frac{R_\mathrm{HMNS}}{v_A}=\frac{R_\mathrm{HMNS}\sqrt{4 \pi \rho}}{B}
    \label{eq:braking_time}
\end{equation}
where $R_\mathrm{HMNS}$ is the radius of the remnant and $v_A$ is the Alfv\'{e}n velocity, the field strength required to activate braking is lower, reflected in the lower saturation value of $E_\mathrm{mag} = \SI{2.57e51}{erg}$ for the \pp case compared to $E_\mathrm{mag} = \SI{7.36e51}{erg}$ reached in the \nsp case. The \mm case remains far from saturation at $\SI{30}{ms}$ postmerger, with magnetic energy growth outpacing the rate inferred from winding alone, suggesting that additional turbulent effects from remnant oscillations or the \mri contribute a significant fraction of the total $E_\mathrm{mag}$.

\subsubsection{Magnetorotational Instability}

\begin{figure*}
    \centering
    \includegraphics[width=0.98\textwidth]{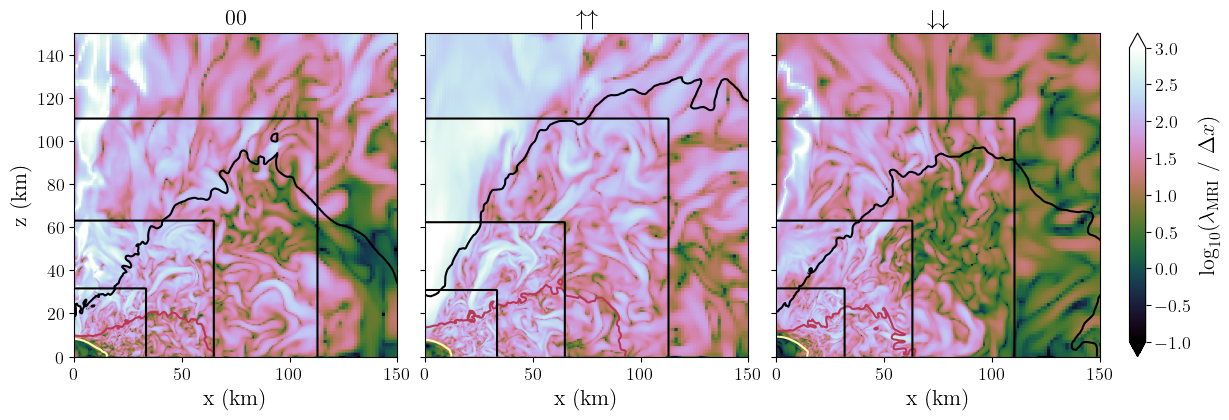}
    \caption{\mri quality factor plotted in the x-z plane at $\SI{30}{ms}$ postmerger. We additionally show the extent of the three finest mesh refinement levels given by the black boxes, and density contours at $10^9, 10^{11},$ and $10^{13}~\mathrm{g~cm}^{-3}$.}
    \label{fig:QMRI}
\end{figure*}

To estimate the activity of the \mri, we calculate the approximate local wavelength of the fastest growing mode as
\begin{equation}
    \lambda_\mathrm{MRI} \approx \frac{2 \pi v_A}{\Omega} = \frac{2 \pi}{\Omega} \frac{|B_p|}{\sqrt{4\pi\rho}}
    \label{eq:lambdaMRI}
\end{equation}
where the subscript $p$ indicates a poloidal component, which holds for a Keplerian disk in Newtonian gravity. In Fig.~\ref{fig:QMRI}, we plot the \mri quality factor $\lambda_\mathrm{MRI} / \Delta x$, the number of grid cells resolving this wavelength, in the x-z plane at $\SI{30}{ms}$ postmerger. As the \mri is typically considered to be partially resolved if $\lambda_\mathrm{MRI} / \Delta x \gtrsim 10$ \cite{siegel_three-dimensional_2018}, our simulations suggest \mri growth is present in the accretion disks within a radius of $\sim \SI{100}{km}$ in all cases, with the strongest activity in the upper layers of the disk at the interface with the polar outflow. The \mri is best resolved in the \pp case, consistent with $\beta^{-1}$ in the left column, middle row of Fig.~\ref{fig:allxz} showing collisions between dynamical and tidal ejecta initializing stronger magnetic fields in the upper layers of the disk, and thus larger $\lambda_\mathrm{MRI}$.

\begin{figure}
    \centering
    \includegraphics[width=0.99\linewidth]{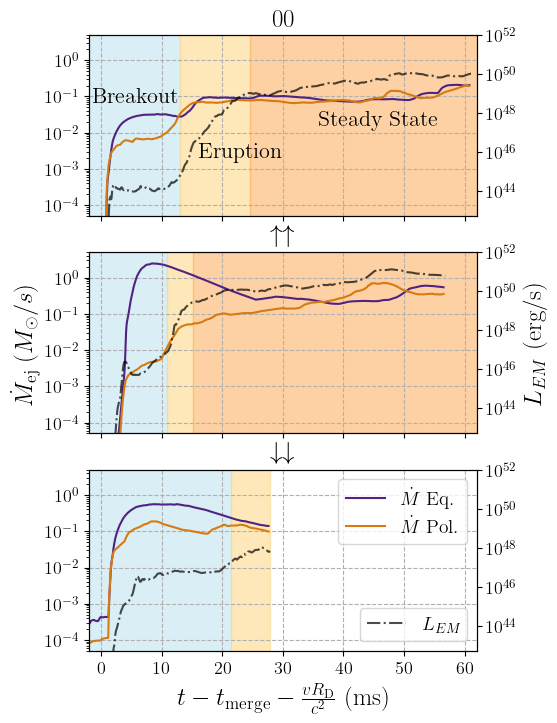}
    \caption{Evolution of mass ejection rate divided into equatorial $(\theta \geq 45^\circ)$ and polar $(\theta < 45^\circ)$ components, along with Poynting luminosity, all extracted from a spherical detector at $R_D = \SI{740}{km}$. We use the Bernoulli criterion (Eq.~\ref{eq:bernoulli}) to flag unbound material. We shift the data by the polar outflow propagation time to the detector surface. The top, middle, and bottom plots show data from the \nspc, \ppc, and \mm simulations, respectively. Following descriptions in Section \ref{sssec:eruption}, the blue, yellow, and orange shaded regions correspond to the breakout, eruption, and steady state phases, as labeled in the upper plot.}
    \label{fig:Mdot_L}
\end{figure}

\subsubsection{Magnetic Eruption}
\label{sssec:eruption}

\begin{figure*}
    \centering
    \includegraphics[width=\textwidth]{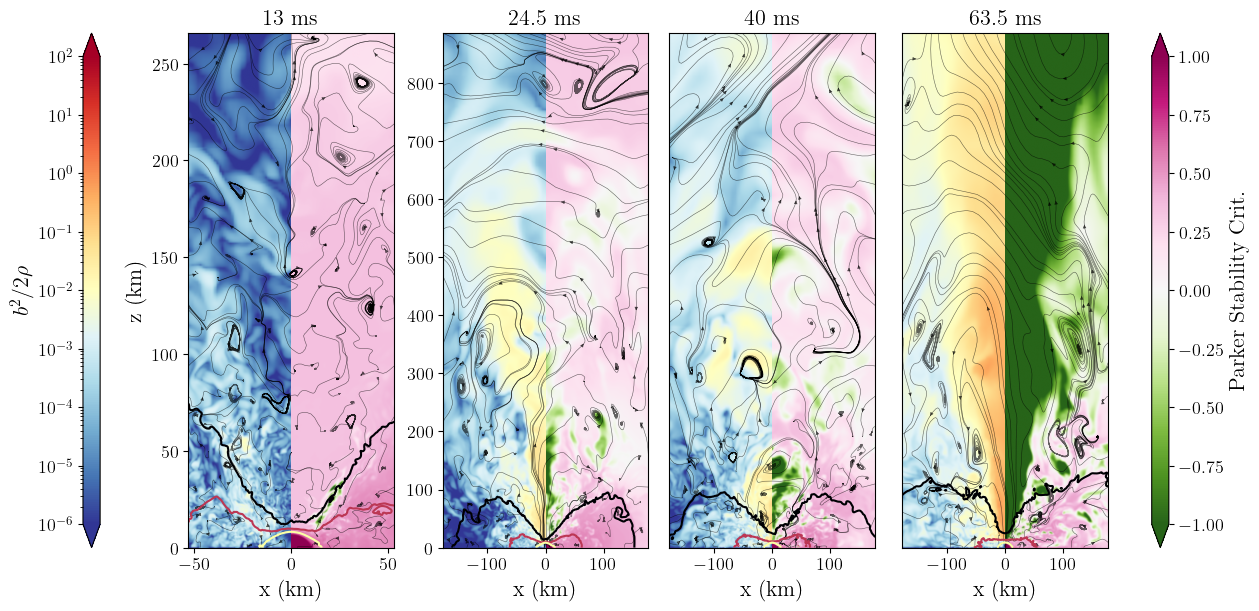}
    \includegraphics[width=\textwidth]{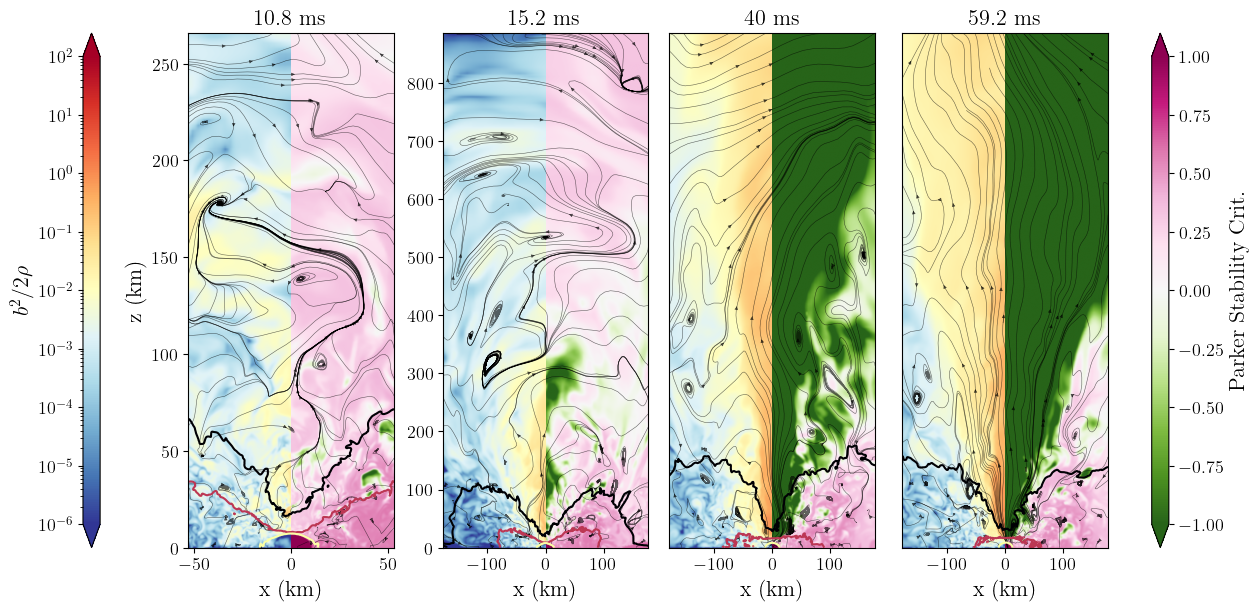}
    \caption{Quantities describing magnetic breakout and propagation in the x-z plane from the \nsp (top) and \pp (bottom) simulations. Panels show characteristic times in the outflow evolution from Fig.~\ref{fig:Mdot_L}, corresponding to, from left to right, the beginning of the eruption phase, the beginning of the steady state, a representative time within the steady state, and the final snapshot. We plot $b^2/2\rho$ on the left of each panel, and the Parker stability criterion from Eq.~(\ref{eq:parker}) on the right. Note that we show a zoomed in view of the remnant in the leftmost plot. We superimpose poloidal magnetic field streamlines and density contours at $10^{13}, 10^{11},$ and $10^9~\mathrm{{g~cm}^{-3}}$.}
    \label{fig:jets}
\end{figure*}

\begin{figure}
    \centering
    \includegraphics[width=\linewidth]{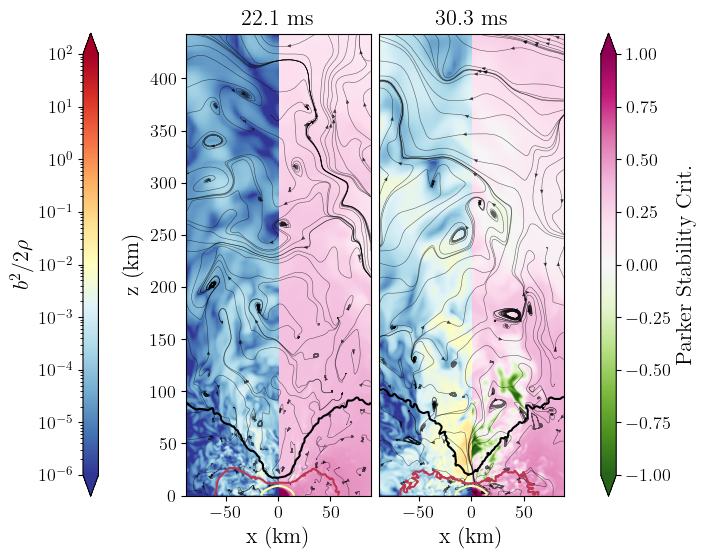}
    \caption{Same as Fig.~\ref{fig:jets} for the \mm case. We only show two panels corresponding to the beginning of the eruption phase, and the final time. Note the different axes extent compared to Fig.~\ref{fig:jets}.}
    \label{fig:jet_mm}
\end{figure}

Following the analysis of \cite{musolino_impact_2025, jiang_long-term_2025}, we observe the growth of magnetic fields to trigger the \parker \cite{parker_dynamical_1966} in the polar funnel of the accretion disk and mediate the transition between distinct postmerger phases. These phases are indicated by the shaded background in Fig.~\ref{fig:Mdot_L}, which also shows the simultaneous response of mass ejection rate $\dot{M}_\mathrm{ej}$ and Poynting luminosity $L_\mathrm{EM}$. Weak magnetic fields first emerge from the remnant in the initial \textit{breakout} phase shaded in blue. The second \textit{eruption} phase, shaded in yellow, coincides with the rapid increase of magnetic pressure in the funnel and the activation of the \parker. In the \nsp and \pp cases, the polar $\dot{M}_\mathrm{ej}$ increases by a factor of $\sim 10$, and $L_{EM}$ increases by a factor of $\gtrsim 100$, signaling the activation of a magnetically driven outflow along the polar axis. This brings the system to a late time \textit{steady state}, shaded in orange, where the magnetic wind stabilizes. In the \nsp case, the transition to eruption and steady-state phases begin at $13$ and $\SI{24.5}{ms}$ postmerger, respectively. This process is accelerated and magnified in the \pp case due to decreased pollution of the polar funnel described in Section \ref{sec:overview}, as the transitions occur at $10.8$ and $\SI{15.2}{ms}$ instead, and the final polar $\dot{M}_\mathrm{ej}$ and $L_\mathrm{EM}$ are higher. Note that in the \mm case, the steady state is never reached, and while we use the appearance of small \parker-unstable regions in the funnel (see below and Fig.~\ref{fig:jet_mm}) to mark the onset of the ``eruption" at $\SI{22.1}{ms}$, we do not see a strong impact on $\dot{M}_\mathrm{ej}$ or $L_\mathrm{EM}$. 

The balance between fluid and magnetic pressures is described by the Parker stability criterion $\mathcal{P} \geq 0 $, where
\begin{equation}
    \mathcal{P} = \frac{d\log P}{d\log \rho} - 1 - \frac{\beta^{-1}(1 + 2 \beta^{-1})}{2 + 3 \beta^{-1}}
    \label{eq:parker}
\end{equation}
such that $\mathcal{P} < 0$ indicates an active instability and magnetic buoyancy in a fluid element. To link the activation of the \parker near the remnant to the eruption of $\dot{M}_\mathrm{ej}$ and $L_\mathrm{EM}$ measured at a spherical detector with radius $R_D = \SI{740}{km}$, in the top row of Fig.~\ref{fig:jets} we plot, for the \nsp case, $\mathcal{P}$ on the right alongside the magnetic energy density to rest-mass density ratio $b^2/2\rho$ on the left. Overlaid streamlines show the structure of the poloidal magnetic field. We choose four snapshots corresponding to the beginning of the eruption phase, the beginning of the steady state, a representative state at $\SI{40}{ms}$ postmerger, and the final snapshot. \parker unstable regions first appear in the disk-funnel boundary at the onset of eruption, where the \mri was best resolved in Fig.~\ref{fig:QMRI}. As the eruption and steady state progress, the \parker extends to the polar cap of the remnant and rises throughout the polar funnel, as was seen in \cite{jiang_long-term_2025}. 

The eruption dynamics in the \pp case shown in the bottom row of Fig.~\ref{fig:jets} agree with the more dramatic transition in $\dot{M}_\mathrm{ej}$ and $L_\mathrm{EM}$. The \parker is only intermittently active in the \nsp case until the funnel becomes fully buoyant $\SI{50}{ms}$ postmerger, while the \pp case shows widespread \parker activity and clear structure in the poloidal field soon after the steady state is reached. The same quantities are shown for the \mm case in Fig.~\ref{fig:jet_mm}, further demonstrating the failure to create a large, organized magnetic outflow within the simulated duration despite starting from similar values of $b^2/2\rho$ and $\beta^{-1}$ (see Fig.~\ref{fig:allxz}) in the funnel as the \nsp case. In each simulation, $b^2/2\rho \lesssim1$, suggesting that even in the steady state, polar outflows remain far from the force-free regime \cite{ruiz_general_2017}.

\begin{figure}
    \centering
    \includegraphics[width=\linewidth]{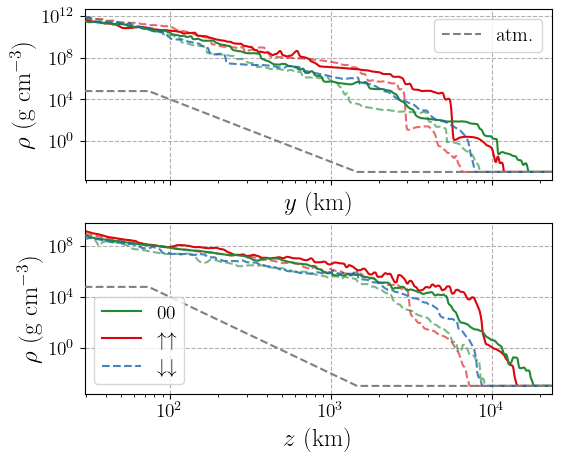}
    \caption{1D profiles of the density $\rho$ along the y and z axes. Dashed colored curves show profiles taken at $\SI{30}{ms}$ postmerger, while solid lines show the final time in \nsp and \pp cases. The gray dashed curve shows the value of the radially decreasing atmosphere floor described in Section \ref{sssec:atm}.}
    \label{fig:1D_rhos}
\end{figure}

In Fig.~\ref{fig:1D_rhos}, we show 1D density profiles along the y and z axes over large distances at $30$ and $\SI{60}{ms}$ postmerger. First, this illustrates the polluted environment in the \mm case. The dynamical ejecta, shown by the outermost front of the density profiles, is densest there, contributing greater resistance to magnetic breakout before $\SI{30}{ms}$ postmerger. This agrees with high values of polar $\dot{M}_\mathrm{ej}$ pre-eruption in Fig.~\ref{fig:Mdot_L}. Exacerbated by ineffective magnetic amplification from the \khi, magnetic terms in Eq.~\ref{eq:parker} are unable to overpower the fluid terms over large regions of the funnel. This leaves the outflow with high densities, comparable to the magnetically driven wind in the \nsp case, but relatively low velocities. When comparing late-time polar outflows, shown from $z\sim10^2 - \SI{8e3}{km}$ in the bottom panel of Fig.~\ref{fig:1D_rhos}, inside the dynamical ejecta front, the density in the \pp outflow is higher than the \nsp case, as expected from higher $\dot{M}_\mathrm{ej}$.

We note the importance of the graded atmosphere floor described in Section \ref{sssec:atm}, shown as gray dashed lines below the density profiles from Fig.~\ref{fig:1D_rhos}. Out to very large distances of $\sim 10^4$ km, the dynamical ejecta front is preserved without significant resistance from the outflow-atmosphere interface, nor atmosphere resets of low density regions.

\begin{figure}
    \centering
    \includegraphics[width=\linewidth]{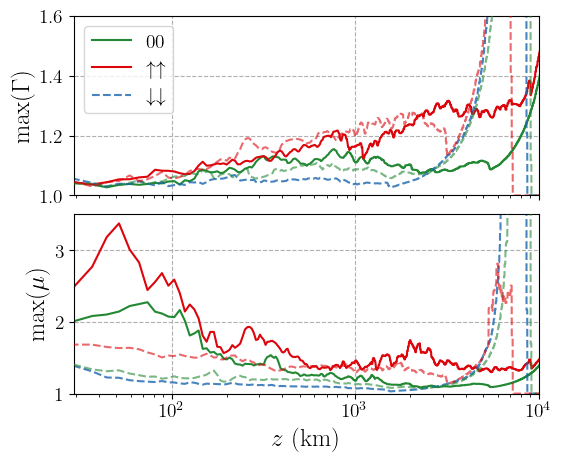}
    \caption{1D profiles of the Lorentz factor $\Gamma$ and energy-to-mass flux ratio $\mu$ from Eq.~(\ref{eq:mu}). We take maximum values of these quantities inside the polar funnel with $\theta < 45^\circ$ at each $z$. As was done in Fig.~\ref{fig:1D_rhos}, dashed curves show profiles taken at $\SI{30}{ms}$ postmerger, while solid curves show the final time in \nsp and \pp cases.}
    \label{fig:1D_gammas}
\end{figure}

We find the polar outflow in the two magnetic wind-producing simulations, \nsp and \ppc, to be dense and mildly relativistic, but inconsistent with typical \sgrb[s] at the end of their respective simulations. As \sgrb jets require terminal Lorentz factors of $\Gamma_\infty > 10$ \cite{piran_physics_2004}, we estimate this value obtained in our simulations with $\Gamma_\infty \sim \mu$, where $\mu$ is the ratio of energy to rest mass flux conserved along poloidal field lines:
\begin{equation}
\begin{split}
    \mu &= \frac{\Gamma^2\rho h v_p + \frac{1}{4 \pi} |\mathbf{E} \times \mathbf{B}_\phi|}{\Gamma\rho v_p} \\
    &\approx \left(h + \frac{b^2}{\rho}\right) \Gamma
    \label{eq:mu}
\end{split}
\end{equation}
and $h$ is the specific enthalpy \cite{vlahakis_relativistic_2003, bamber_jetlike_2024}. The second approximation assumes the dominant component of the velocity to be poloidal ($v_p \gg v_\phi$) while the magnetic field is toroidal ($B_\phi \gg B_p$). In Fig.~\ref{fig:1D_gammas}, we plot the maximum value of $\Gamma$ and $\mu$ at each $z$ within a polar cone of $\theta < 45^\circ$. While both the simulated and inferred asymptotic Lorentz factors are higher in the \pp case, both \nsp and \pp cases exhibit $\mu \sim O(1)$ in the magnetized outflow, placing strong constraints on early \sgrb emission. Thus, even in the \pp case, where the combined effects of neutrino cooling and a cleaner polar funnel create an ideal environment for magnetic breakout and collimation, the resulting magnetar wind is incompatible with a typical \sgrb jet.

\begin{figure}
    \centering
    \includegraphics[width=\linewidth]{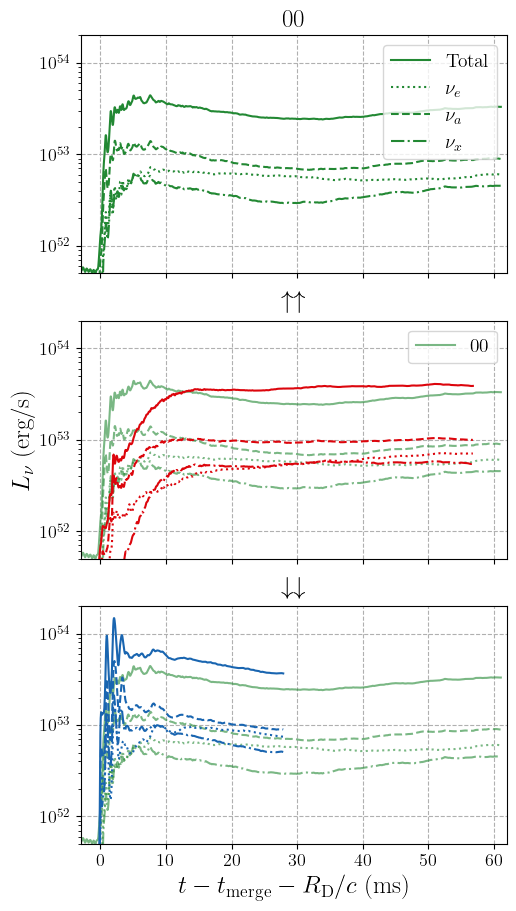}
    \caption{Neutrino luminosities of all evolved neutrino species for the \nsp (top), \pp (middle), and \mm (bottom) cases. Each curve has been shifted by the propagation time to the detector at $R_D = \SI{740}{km}$. Data from the \nsp case is shown in the background in other panels to provide a comparison. Note that the plotted $L_{\nu_x}$ shows the luminosity of one representative heavy neutrino type, such that the heavy neutrino contribution to the total luminosity is $4L_{\nu_x}$.}
    \label{fig:Lnu}
\end{figure}

\begin{figure*}
    \centering
    \includegraphics[width=\textwidth]{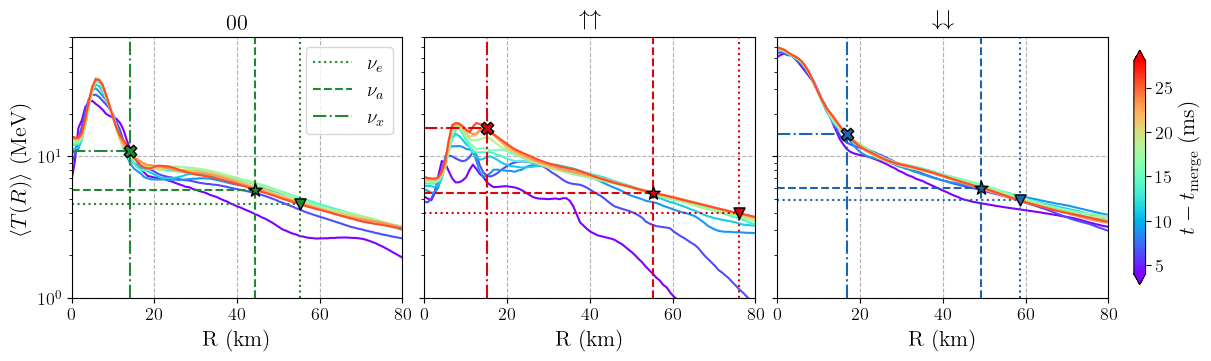}
    \caption{A series of azimuthally averaged temperature profiles for each simulation from 4 to $\SI{28}{ms}$ postmerger, calculated using Eq.~(\ref{eq:phiavg}). The triangle at the intersection of dotted lines in each plot marks the temperature and radius of the electron neutrino surface calculated from Eq.~(\ref{eq:optdepth}). Likewise, the star and dashed lines mark the antielectron neutrino surface, and the X and dash-dotted line marks the heavy neutrino surface.}
    \label{fig:T_Rnu}
\end{figure*}

\subsection{Neutrino Emission}
\label{ssec:neutrino}

The distinct postmerger environments induced by initial spins result in different trends in the neutrino emission responsible for cooling the remnant, and changing the ejecta composition \cite{radice_dynamical_2016, siegel_three-dimensional_2018, mosta_magnetar_2020, zappa_binary_2023}. Figure~\ref{fig:Lnu} shows the luminosity of each neutrino species across spin cases. As neutrino emissivities are highly sensitive to the fluid temperature, the evolution of the luminosity follows $\langle T(R,t)\rangle$ in Fig.~\ref{fig:T_Rnu}. At early times before $\SI{10}{ms}$ postmerger, the luminosity is set by the heat generated by the initial collision at merger, following $L_\nu^{\downarrow\downarrow} > L_\nu^{00} > L_\nu^{\uparrow\uparrow}$ for each neutrino species. This is reflected in the early, purple curves in Fig.~\ref{fig:T_Rnu} for each spin case. As discussed in Section \ref{sssec:khi}, the \nsp case reflects a typical \hmns remnant with a relatively cold core and hot envelope. The \pp case is initially colder, with tidal ejecta remaining below $\sim \SI{1}{MeV}$ in the inner disk. A persistent double peaked temperature profile is apparent between $5-\SI{20}{km}$ in the central panel of Fig.~\ref{fig:T_Rnu}. The \mm case, experiencing the most violent merger, exhibits high temperatures throughout the remnant and into the inner disk that stay relatively constant. However, after $\SI{15}{ms}$, more dynamical ejecta from the \pp merger collides with the dense tidal ejecta, converts kinetic to thermal energy, and falls back onto the remnant, raising the temperature most dramatically in the outer remnant and inner disk ($R > \SI{15}{km}$) where the neutrinosphere surfaces lie. This allows the luminosity across all neutrino species in the \pp case to exceed the \nsp case, and the luminosity for all species except $\nu_e$ to exceed the \mm case at $\SI{30}{ms}$ postmerger. At the end of the simulation, each case satisfies $L_{\nu_a}>L_{\nu_e}>L_{\nu_x}$, following the trend from previous studies using this neutrino transport scheme (e.g., \cite{zappa_binary_2023, radice_ab-initio_2023, espino_impact_2024}).

For a more quantitative connection between the thermodynamics of the remnant and neutrino emission, Fig.~\ref{fig:T_Rnu} also shows radii and temperatures of the neutrinospheres in the x-y plane, marked with vertical and horizontal lines. We estimate the location of the neutrinosphere by finding the radius where the optical depth $\tau_{\nu_i}=1$, using the azimuthally averaged equilibration opacity $\langle \kappa_{\mathrm{eq},\,\nu_i}\rangle$ \cite{endrizzi_thermodynamics_2020}:
\begin{equation}
\begin{split}
    \tau_{\nu_i}(R_{\nu_i}) & = \int_{R_\mathrm{max}}^{R_{\nu_i}} \langle \kappa_{\mathrm{eq},\,\nu_i}\rangle \,dr \\
    & =\int_{R_\mathrm{max}}^{R_{\nu_i}} \left\langle \sqrt{\kappa_{a,\,\nu_i} (\kappa_{a,\,\nu_i} + \kappa_{s,\,\nu_i})}\right\rangle\,dr
    \label{eq:optdepth}
\end{split}
\end{equation}
where $\kappa_{a,\,\nu_i}$ is the absorption opacity for a neutrino species, and $\kappa_{s,\,\nu_i}$ is the scattering opacity. The term in the integrand approximates the opacity where neutrinos are no longer in thermal or weak equilibrium with the fluid at the neutrinosphere. Each simulation satisfies $R_{\nu_e} > R_{\nu_a} > R_{\nu_x}$ and $T(R_{\nu_e}) < T(R_{\nu_a}) < T(R_{\nu_x})$ in the x-y plane, a robust result across previous studies employing different prescriptions for neutrino energies and opacities \cite{endrizzi_thermodynamics_2020, chiesa_open-source_2025}. Following the hierarchy of heavy neutrino luminosities, $T^{\uparrow\uparrow}(R_{\nu_x}) > T^{\downarrow\downarrow}(R_{\nu_x}) > T^{00}(R_{\nu_x})$, and following luminosities across species, $T^{\downarrow\downarrow}(R_{\nu_i}) > T^{00}(R_{\nu_i})$. However, the \pp neutrino surface temperature at $R_{\nu_e}$ and $R_{\nu_a}$ is lowest across all cases despite their luminosities exceeding the \nsp case, suggesting that asymmetric radiation out of the x-y plane is significant here.

\subsection{Ejecta Properties}
\label{ssec:ejecta}

\begin{figure*}
    \centering
    \includegraphics[width=\textwidth]{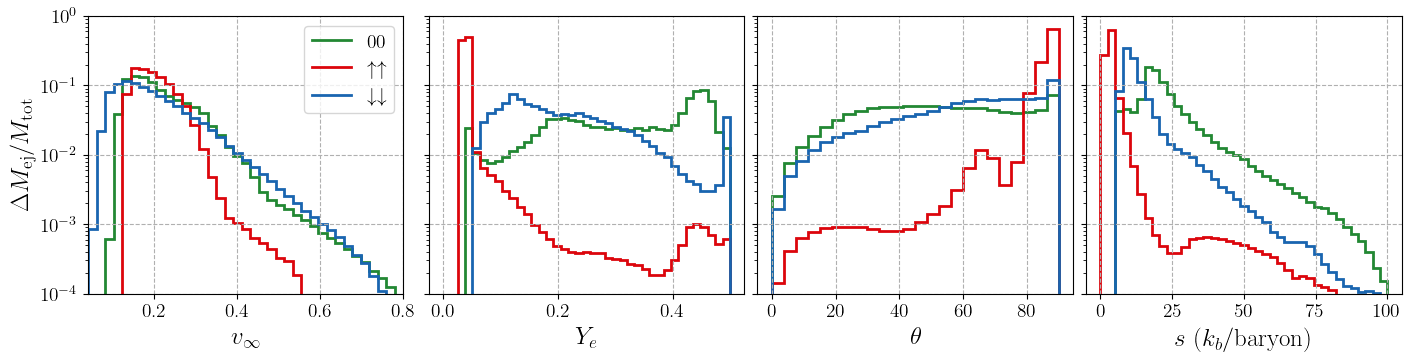}
    \caption{Histograms describing the distribution of dynamical ejecta over, from left to right, terminal velocity $v_\infty$, electron fraction $Y_e$, polar angle $\theta$, and specific entropy $s$, for each spin case. We flag unbound dynamical ejecta at the detector surface by considering material that satisfied the Bernoulli criterion (Eq.~(\ref{eq:bernoulli})) when crossing the surface before the steady state identified in Fig.~\ref{fig:Mdot_L}.}
    \label{fig:hist_dyn}
\end{figure*}

\begin{figure*}
    \centering
    \includegraphics[width=\textwidth]{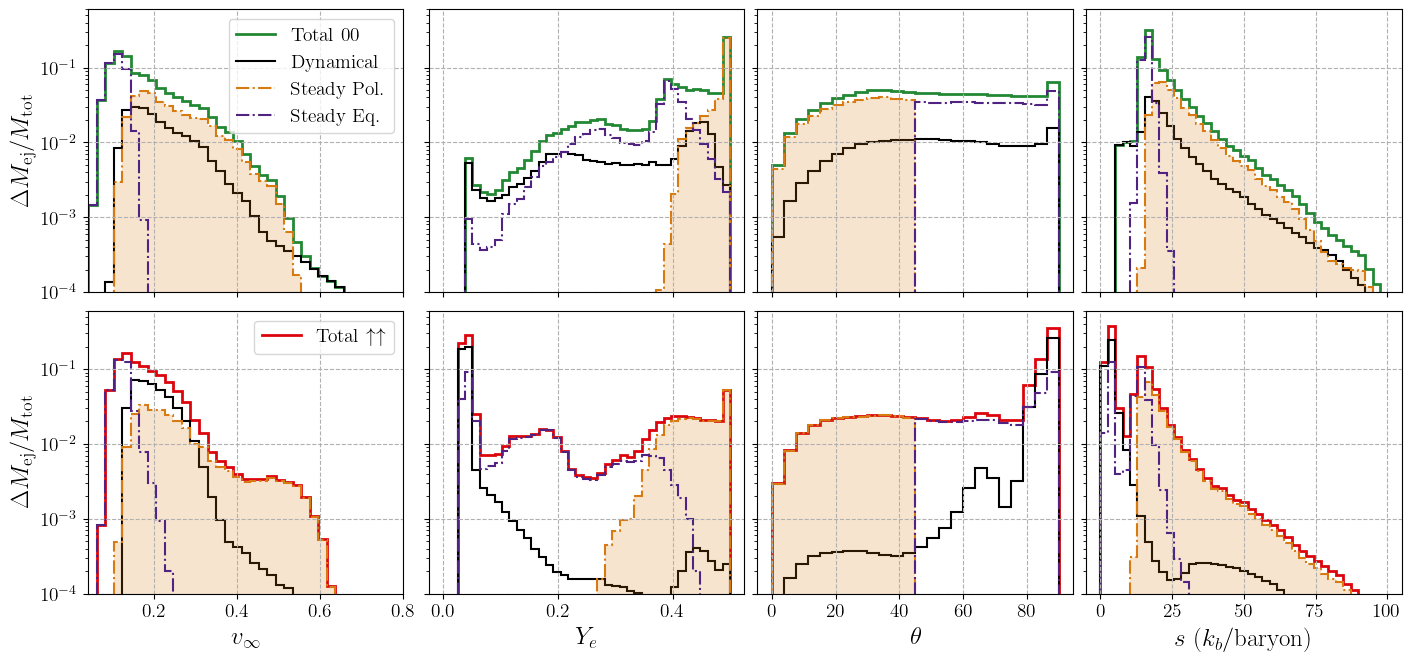}
    \caption{Histograms describing the contribution of various ejecta components to the total mass distribution in the \nsp (top) and \pp (bottom) simulations. The black histogram shows the dynamical ejecta from Fig.~\ref{fig:hist_dyn}. The purple and shaded orange histograms show the equatorial ($\theta \geq 45^\circ$) and polar ($\theta < 45^\circ$) contributions to the steady state ejecta.}
    \label{fig:hists_ss}
\end{figure*}

Thermodynamic properties of unbound material have critical effects on the nucleosynthetic yield and \kNe signal estimated from \bns merger simulations. Notably, ejecta with $Y_e \lesssim 0.25$ are responsible for creating elements beyond the second r-process peak \cite{radice_dynamical_2016}, and quantifying the amount of neutron-rich ejecta provides an estimate of the relative contribution of \bns mergers towards observed r-process abundances. Simulations employing modern M1 schemes have also noted the importance of proton-rich ($Y_e \sim 0.5$) ejecta \cite{bernuzzi_long-lived_2025, neuweiler_general-relativistic_2026}, which could efficiently produce and fuse $\alpha$ particles to build $Z=N$ nuclei up to $^{56}\mathrm{Ni}$ \cite{darbha_nickel-rich_2010}. A recent study evolved the outflow from unmagnetized \bns merger simulations employing the \thcm neutrino transport module to 100 days postmerger using a radiation hydrodynamics code coupled to live nuclear network evolution \cite{magistrelli_element_2024, jacobi_56ni_2026}. When considering the neutrino-driven wind from a long-lived remnant, they found $\alpha$ particles, $^{56}\mathrm{Ni}$, and $^{56}\mathrm{Co}$ to be the dominant nucleosynthesis products, whose subsequent decays after $\sim10-100$ days significantly altered the nuclear heating rate and \kNe light curve, potentially providing a powerful tool to estimate the remnant's lifetime.

We first present a standardized comparison across all spin cases through histograms describing the dynamical ejecta mass distribution across terminal velocity $v_\infty$ (see Eq.~(\ref{eq:vinf})), $Y_e$, polar angle $\theta$, and specific entropy $s$ in Fig.~\ref{fig:hist_dyn}. We define dynamical ejecta as material ejected during breakout or eruption phases present in all simulations and shaded in blue and yellow in Fig.~\ref{fig:Mdot_L}, before the late-time steady state. The \nsp case is broadly consistent with existing literature on dynamical ejecta modeled with an M1 scheme \cite{zappa_binary_2023, espino_impact_2024, schianchi_m1_2023, daszuta_gr-athena_2026}. Histograms show a roughly isotropic distribution, a tail to high $v_\infty$, a large portion of neutron-rich $Y_e$ with the beginnings of a neutron-poor neutrino-driven wind, and shock-reprocessed high values of $s$. Tidal ejecta comprises the majority of the \pp histograms, visible in sharp peaks of low $Y_e$, equatorial $\theta$, and low $s$ material. Dynamical ejecta velocity is lowest here, in agreement with how the leading front of the density profiles in Fig.~\ref{fig:1D_rhos} travels the slowest. The violent collision in the \mm case results in more ejecta with $v_\infty$ from $\sim 0.3 - 0.7$ compared to the \nsp case, along with lower $Y_e$ and $s$ material ejected directly from the initial \ns[s] without additional shock heating or neutrino reprocessing.

\begin{figure*}
    \centering
    \includegraphics[width=0.8\textwidth]{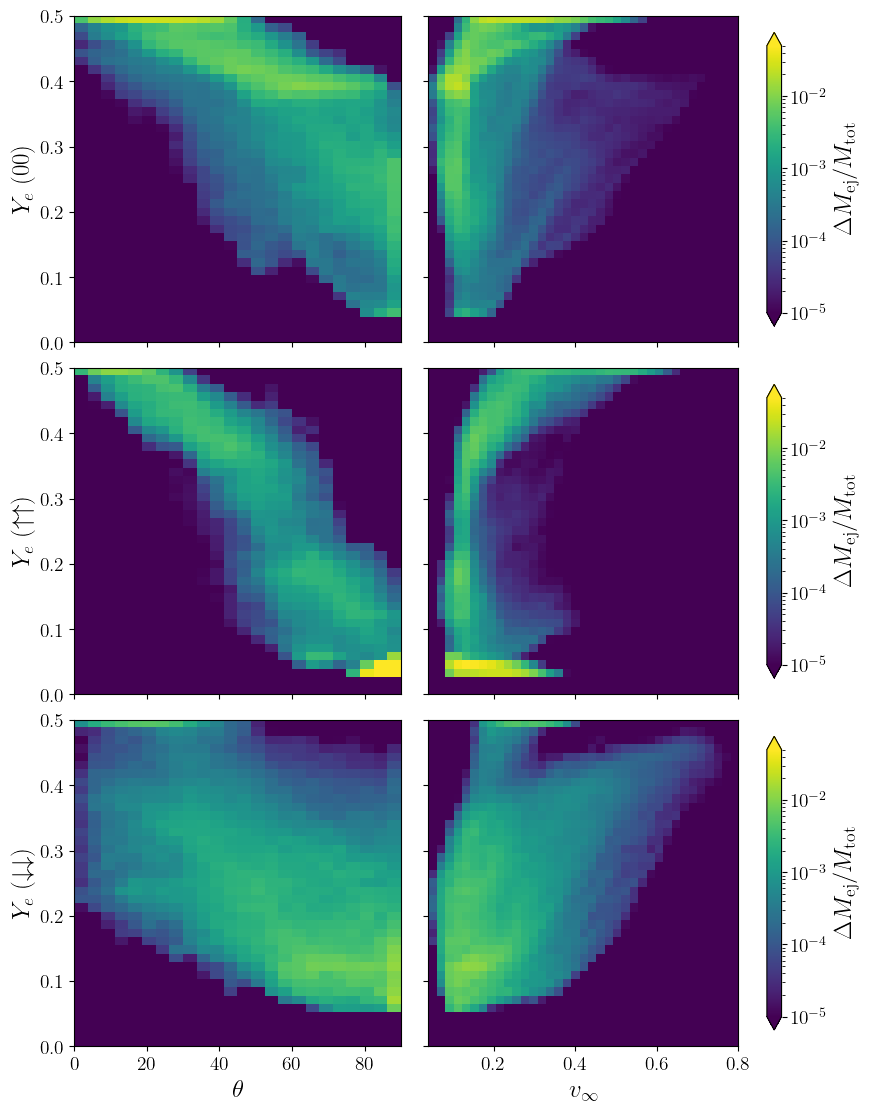}
    \caption{2D histograms showing the ejecta mass distributions over electron fraction $Y_e$ and polar angle $\theta$ in the left column, and $Y_e$ and terminal velocity $v_\infty$ on the right. We show data from the \nspc, \ppc, and \mm simulations in the top, bottom, and middle rows.}
    \label{fig:hist2d}
\end{figure*}

The combined effect of magnetic eruption and prolonged neutrino absorption has a strong effect on the late-time polar ejecta in the \nsp and \pp cases. Recall that Fig.~\ref{fig:jets} shows the \parker unstable region to cover the entire neutrino irradiated funnel, allowing buoyant magnetic fields to carry heavily reprocessed, high $Y_e$ material from near the neutrino surfaces to large distances. The top row of Fig.~\ref{fig:hists_ss} shows the full breakdown of unbound material in the \nsp simulation across dynamical ejecta (i.e. the distribution shown in Fig.~\ref{fig:hist_dyn}), and steady state ejecta divided between polar ($\theta < 45^\circ$) and equatorial ($\theta \geq 45^\circ$) contributions. The polar ejecta, shown in shaded orange, has $Y_e$ values almost entirely above $0.4$. It also features a wide spread of mildly relativistic $v_\infty$ and a tail to high $s$, unlike the steady state equatorial ejecta, which has $s$ over a narrower distribution and low $v_\infty$. Nevertheless, the steady state equatorial ejecta is proton-rich as well, with a dominant peak at $Y_e \approx 0.4$.  We caution that slow velocities in the equatorial ejecta makes this component highly sensitive to simulation duration, as short $O(\SI{100}{ms})$ simulations fail to capture neutron rich disk winds ejected over $\sim\SI{1}{s}$ postmerger due to viscous angular momentum transport \cite{sprouse_emergent_2024, fujibayashi_mass_2020, kawaguchi_long-term_2025}. Similar patterns are visible in the \pp mass distribution in the bottom row of Fig.~\ref{fig:hists_ss}, showing that when heavy magnetar outflows appear in \grmhd simulations coupled to a classical M1 scheme, the proton-richness of the outflow is a robust prediction even when considering an extreme spin case. However, we note that improved neutrino microphysics via direct sampling of the neutrino distribution function in a Monte Carlo scheme \cite{foucart_robustness_2024}, inclusion of muon neutrino and antineutrino species \cite{ng_accurate_2025}, or approximate neutrino flavor oscillations \cite{qiu_neutrino_2025, qiu_impact_2025, lund_angle-dependent_2025} can all shift the ejecta distribution towards lower $Y_e$, especially in polar material ejected over secular timescales. Future \grmhd simulations with increasingly accurate microphysics will be required to clarify the chemical composition of this outflow.

\begin{table}
    \centering
    \setlength{\tabcolsep}{0.8em} 
    {\renewcommand{\arraystretch}{1.25}
    \begin{tabular}{l|c c c}
         Spin Config. & \nsp & \pp & \mm\\
         \hline
         Eq. $M_\mathrm{ej}~[10^{-3}M_\odot]$  & 5.03 & 33.9 & 9.23 \\
         $\langle Y_e\rangle_\mathrm{Eq.}$ & 0.32 & 0.09 & 0.17 \\
         $\langle v_\infty \rangle_\mathrm{Eq.}~[c]$ & 0.13 & 0.16 & 0.19 \\
         $\langle S\rangle_\mathrm{Eq.}~[k_b/\mathrm{baryon}]$ & 16.2 & 6.6 & 10.9 \\
         \hline
         Pol. $M_\mathrm{ej}~[10^{-3}M_\odot]$ & 4.37 & 11.4 & 3.11 \\
         $\langle Y_e\rangle_\mathrm{Pol.}$ & 0.47 & 0.44 & 0.32 \\
         $\langle v_\infty \rangle_\mathrm{Pol.}~[c]$ & 0.25 & 0.26 & 0.23 \\
         $\langle S\rangle_\mathrm{Pol.}~[k_b/\mathrm{baryon}]$ & 29.2 & 22.3 & 20.6 \\
    \end{tabular}
    }
    \caption{In this table, we list the total ejected mass $M_\mathrm{ej}$ determined via the Bernoulli criterion (Eq.~(\ref{eq:bernoulli})), and mass-weighted averages (Eq.~(\ref{eq:mass_avg})) of the electron fraction $\langle Y_e\rangle$, terminal velocity $\langle v_\infty\rangle$, and specific entropy $\langle S\rangle$ split between equatorial ($\theta \geq 45^\circ$) and polar ($\theta< 45^\circ$) components, for each spin configuration.}
    \label{tab:all_ejecta}
\end{table}

In Fig.~\ref{fig:hist2d}, we show 2D ejecta mass distributions over $Y_e$ and either $v_\infty$ or $\theta$ for all unbound material. In the \nsp case shown in the top row, the distribution is broadly consistent with a two-component \kNe model. $Y_e$ is negatively correlated with $\theta$, and higher $Y_e$ ejecta shows a longer tail to high velocities. The diffuse high $v_\infty$ tail is associated with the dynamical ejecta, while the more massive core of the distribution describes the mildly relativistic magnetic wind that extends to high $Y_e$ and $v_\infty \sim 0.5$. Thus, high $Y_e$ material is generally ejected faster at polar angles, leading to less production of opaque heavy elements that serve to redden a \kNe signal \cite{banerjee_opacity_2022}. However, the full cluster of material with $Y_e > 0.38$ extends to angles as high as $80^\circ$, with a distinct component at $Y_e \sim 0.4$, $50^\circ<\theta<85^\circ$, and $v_\infty < 0.15$. The appearance of this component is supported by waves of $Y_e \sim 0.4$ material in the disk shown in the upper right panel of Fig.~\ref{fig:allxz}, and the 1D $Y_e$ histogram of late time equatorial ejecta in the top row, second panel of Fig.~\ref{fig:hists_ss} showing a peak at $0.4$. This component does not appear in either of the spinning cases, suggesting its origin derives from features unique to the \nsp case, including mixing between a dense, but poorly collimated, proton-rich polar outflow and the disk, and greater neutrino reabsorption due to lower optical depth between the remnant and disk inferred from $\rho(y)$ profiles in Fig.~\ref{fig:1D_rhos} and a smaller $\nu_e$ neutrinosphere radius from Fig.~\ref{fig:T_Rnu}. The distinction between polar and equatorial clusters is more explicit in the \pp case, where the ejecta is visibly grouped into two populations of high $Y_e$, low $\theta$, and $v_\infty$ extending up to $\sim0.75$, or low $Y_e$, high $\theta$, and low $v_\infty$. In contrast, the distributions are less clustered in the \mm case, reflecting more isotropic dynamical ejecta and the lack of a collimated outflow with distinct properties. We summarize total ejecta mass and mass-weighted averages divided between equatorial and polar components in Table \ref{tab:all_ejecta} and leave direct calculations of \kNe light curves to a future study.

Table~\ref{tab:all_ejecta} shows that the spinning cases eject more total mass than the \nsp case. This is the result of both orbital dynamics described in Sec.~\ref{sec:overview} and magnetically driven outflows described in Sec.~\ref{sssec:eruption}, and follows the mass ejection rate in Fig.~\ref{fig:Mdot_L}. In the \pp case, $\dot{M}_\mathrm{ej}$ shows a massive contribution from slow tidal ejecta after the disruption of the stars immediately prior to merger (see also density contours in central row, left column of Fig.~\ref{fig:allxz} and discussion of Fig.~\ref{fig:hist_dyn}). Note that this component is shown to appear after $t-t_\mathrm{merge} = 0\,\mathrm{ms}$ in Fig.~\ref{fig:Mdot_L} as data was shifted by the polar outflow propagation time while tidal ejecta is much slower. Combined with a more efficient polar outflow in the steady state phase, the \pp case ejects the most material in both equatorial and polar directions. In the \mm case, it is instead dynamical ejecta from a more violent merger that provides the mechanism to release more equatorial ejecta than the \nsp case, but the lack of a magnetically driven outflow and a shorter simulation limits polar mass ejection.

We caution that $Y_e$ values were erroneously capped at $0.5$ in the simulation domain, causing a visible pile-up of material in the highest $Y_e$ bin in each histogram. While we do not expect this to affect the overall dynamics of \mhd, this does suppress information on the distribution of proton-rich ejecta and biases the calculation of mass-averaged quantities in the polar outflow.

\subsection{Nucleosynthesis}
\label{ssec:nuc}

\begin{figure}
    \centering
    \includegraphics[width=\linewidth]{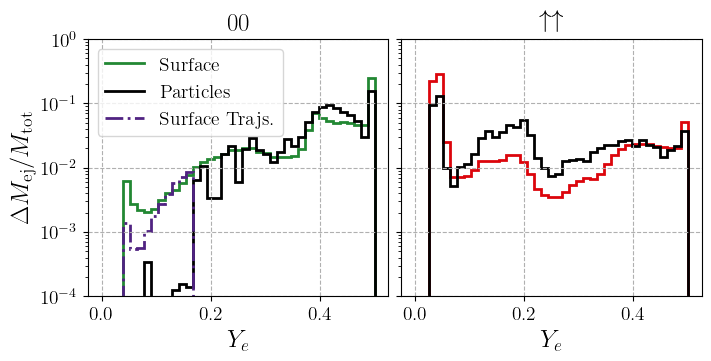}
    \caption{Mass distributions over $Y_e$ for unbound fluid measured at a spherical detector with $R_D = \SI{740}{m}$ compared to unbound tracer particles, shown as colored and black histograms, respectively, for \nsp (left) and \pp (right) simulations. For the \nsp case, we additionally plot the contribution of binned trajectories derived from a smaller spherical detector at $\SI{295}{km}$ in purple. These are used to supplement the reaction network calculation as described in Section \ref{ssec:nuc}.}
    \label{fig:outflow_pt}
\end{figure}

\begin{table}
    \centering
    \setlength{\tabcolsep}{0.8em} 
    {\renewcommand{\arraystretch}{1.6}
    \begin{tabular}{l|c c c}
         \# tracers & \nsp & \pp & \mm\\
         \hline
         Total & 1556 & 5616 & 1728 \\
         Polar & 443 & 957 & 521 \\
    \end{tabular}
    }
    \caption{The number of tracer particles used for nucleosynthesis calculations in each simulation. Particles are first flagged as unbound if they satisfy the Bernoulli criterion at the end of the simulation (Eq.~\ref{eq:bernoulli}), and from that subset, classified as polar if their location at the end of the simulation lies within a polar angle of $\theta < 45^\circ$.}
    \label{tab:ntracers}
\end{table}

We postprocess simulation data to estimate nucleosynthesis yields by providing particle tracer trajectories to the \skynet nuclear reaction network as described in Appendix \ref{ax:analysis}. We first remark on our ability to capture outflow characteristics using tracer particles. Table \ref{tab:ntracers} lists the number of unbound tracers used for nucleosynthesis calculations. The total number exceeds $1500$ in each case, with over $400$ resolving the polar outflow, demonstrating adequate sampling of the domain. Figure~\ref{fig:outflow_pt} compares the distribution of $Y_e$ over unbound tracers compared to the distribution measured at a spherical detector for \nsp and \pp simulations (i.e. Fig.~\ref{fig:hists_ss}). In the \nsp case, the tracers effectively represent the distribution from $Y_e$ of $0.17-0.5$, including the high $Y_e$ steady state polar outflow, but undersample low $Y_e$ dynamical ejecta. To recover this component, we apply the method from \cite{radice_dynamical_2016} to construct ejecta trajectories from spherical detector data, that is, record $\rho_0,\,s_0,$ and $Y_{e,\,0}$ of unbound fluid elements at a detector of radius $\SI{295}{km}$, assume the density follows homologous expansion with a timescale $\tau_\mathrm{ex}$ such that
\begin{equation}
    \rho(t) = \rho_0 \left( \frac{v_r}{R}t\right)^{-3} \equiv \rho_0 \left( \frac{3\tau_\mathrm{ex}}{et}\right)^{3}
    \label{eq:of_traj}
\end{equation}
where $e$ is Euler's number, and bin unbound fluid elements with similar characteristics to assign a mass to the trajectory. We construct trajectories for unbound fluid elements crossing the detector before $\SI{24.5}{ms}$, i.e.~before the late-time steady state, with $Y_e < 0.17$ and show their contribution to the full dataset provided to \skynet as the purple histogram in Fig.~\ref{fig:outflow_pt}. Results from \cite{radice_binary_2018} showed this method to differ from using tracer particle trajectories by up to a factor of two near the first r-process peak but agree within $\sim20\%$ from $140 \leq A \leq 190$. The \pp particles overrepresent the steady state disk ejecta from $Y_e$ of $0.1-0.35$ and underrepresent the tidal component, leaving the overall amount of neutron rich ejecta consistent with the outflow. 

\begin{figure}
    \centering
    \includegraphics[width=\linewidth]{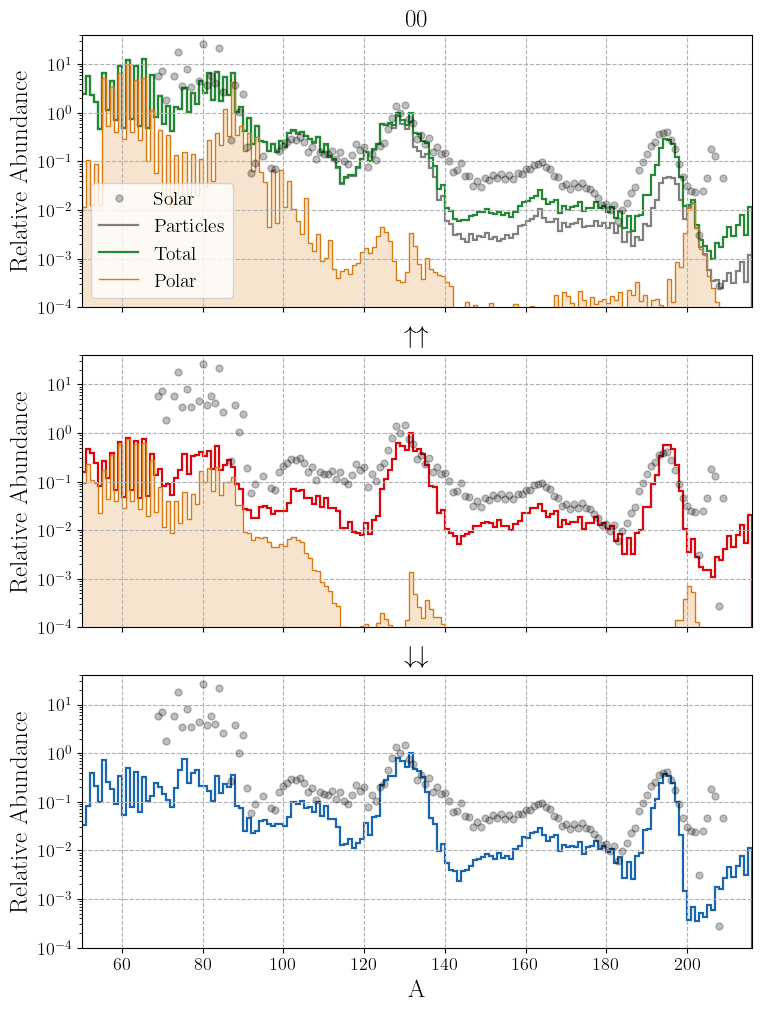}
    \caption{Relative abundances of nucleosynthesis yields after $10^9$ s of \skynet evolution for each simulation. We show inferred solar r-process abundances as gray circles \cite{arnould_r-process_2007}. Total abundances, shown in colored lines, are normalized to the maximum value in the second peak. The \nsp and \pp plots show the contributions from polar tracers in shaded orange bars. Additionally, in the upper \nsp plot, we provide yields from particle tracers alone in gray to show the impact of including trajectories derived from the spherical outflow detector on the total abundance pattern.}
    \label{fig:nuc}
\end{figure}

The resulting r-process yield from \skynet for each simulation, shown in Fig.~\ref{fig:nuc}, complements the neutron richness of the ejecta described in Section \ref{ssec:ejecta}. We observe good agreement with solar abundances in the \nsp case for mass numbers below the second peak. As shown by the difference between the green and gray lines, despite only comprising $3\%$ of the total ejecta mass, the supplemental trajectories derived from the spherical detector are critical for capturing nucleosynthesis beyond the second peak. All cases then recreate the relative abundances in the second and third peaks, consistent with previous M1 studies focusing on dynamical ejecta \cite{radice_new_2022, espino_impact_2024}.

In Fig.~\ref{fig:nuc}, we again highlight the contribution of the polar component in the \nsp and \pp cases in orange. High velocity material in the polar dynamical ejecta shock front contributes a shifted third peak centered around $A \sim 200$ due to slower neutron capture rates building nuclei closer to the region of stable nuclear isotopes \cite{mendoza-temis_nuclear_2015}. Otherwise, the polar outflow only provides a modest contribution to r-process yields, suppressed by a factor of $\sim 10^2-10^3$ relative to the equatorial component over values of $A$ beyond the second peak. 

\begin{table}[ht!]
    \centering
    \setlength{\tabcolsep}{0.8em} 
    {\renewcommand{\arraystretch}{1.2}
    \begin{tabular}{c|c c c}
         $M_\mathrm{ej}$ [$10^{-3}~M_\odot$] & \nsp & \pp & \mm\\
         \hline
         $M_\mathrm{ej}^n$ & 1.37 & 30.95 & 8.66 \\
         $M_\mathrm{ej}^p$ & 2.38 & 2.37 & 0.43 \\
         $\alpha$ & 1.58 & 2.10 & 1.16 \\
         $^{56}\mathrm{Ni}$ & 0.20 & 0.14 & 0.18
    \end{tabular}
    }
    \caption{Ejecta masses of neutron rich $(Y_e \leq0.25)$ material $M_\mathrm{ej}^n$, proton rich $(Y_e \geq 0.49)$ material $M_\mathrm{ej}^p$, $\alpha$ particles, and $^{56}\mathrm{Ni}$ nuclei at $\SI{10}{s}$ postmerger. Values were calculated by multiplying the total ejecta mass in Table \ref{tab:all_ejecta} by the isotopic mass fraction reported by \skynet.}
    \label{tab:ni56_yield}
\end{table}

Instead, we find significant production of first peak ($73 \leq A \leq 91$) and iron group ($50 \leq A \leq 56$) elements. While our artificial limit of $Y_e \leq 0.5$ suppresses the formation of $Z=N$ nuclei by removing excess protons and driving nuclei towards stability, we still observe significant production of $\sim 10^{-4}~M_\odot$ of $^{56}\mathrm{Ni}$ and $\sim 10^{-3}~M_\odot$ of $\alpha$ particles $\SI{10}{s}$ into \skynet evolution, demonstrating that the environment still achieves conditions in $Y_e$, $s$, and expansion velocity compatible with $^{56}\mathrm{Ni}$ production. These light nucleosynthesis yields are listed in Table \ref{tab:ni56_yield}. While the factor of $6$ difference in proton-rich ejecta between magnetically erupting cases and the \mm case highlights the possibility of the outflow enhancing this proposed emission channel, $^{56}\mathrm{Ni}$ production does not exhibit the same contrast across spin cases. Thus, while our results support the conclusion from \cite{jacobi_56ni_2026} that a \kNe light curve consistent with the decay of key iron group elements is a strong indicator of a long-lived remnant where neutrino and magnetic pressures are driving a proton-rich outflow, investigating how different postmerger environments affect precise nucleosynthesis yields remains an important avenue of future study.


\section{Summary and Conclusion}
\label{sec:summary}

We performed three \bns merger simulations with \igm coupled to an M1 neutrino transport scheme, featuring three equal mass spin configurations: a fiducial nonspinning case (\nspc), a case with spins aligned with the orbital angular momentum (\ppc, with $\chi_{NS}=+0.43$), and a case with spins antialigned with the orbit (\mmc, with $\chi_{NS}=-0.43$). Our choices of using the APRLDP \eos model and low initial \ns masses of $M = \SI{1.35}{M_\odot}$ create long-lived remnants that resist collapse in the $\SI{60}{ms}$ of postmerger evolution we simulate, leading to significant neutrino emission and magnetic amplification from the remnant. The resulting interplay of neutrino feedback onto ejecta launched from a magnetic eruption leads to significant protonization and mass-loading of the outflow, with further implications on potential \kNe and \sgrb emission. 

Qualitative dynamics resulting from intrinsic spins in the initial stars affect many aspects of postmerger evolution. Due to spin-orbit coupling, the \mm case inspirals faster than the \nsp case, leading to a more violent collision at merger. This collision expels a heavy cloud of dynamical ejecta and bound debris, immediately disrupts the \khi-unstable shear layer between the stars, suppresses magnetic amplification within the remnant, and impedes the early collimation of magnetic outflows. The violence of the collision also sets high initial temperatures and neutrino luminosities in the remnant, leading to brighter neutrino radiation. The anti-aligned spins of the initial \ns[s] result in lower angular velocities in the postmerger remnant, reducing the $f_1$ and $f_2$ peaks of the postmerger \gw spectrum by $\sim O(\SI{100}{Hz})$.

Conversely, the slower \pp inspiral allows the \ns[s] to tidally disrupt, releasing dense arms of tidal ejecta into the equatorial plane prior to merger, which provides a dominant source of neutron rich ejecta and reduces the amount of dynamically ejected debris in the polar funnel. Subsequent collisions between dynamical and tidal ejecta in the \pp case reheat the remnant, leading to comparable neutrino emission between all spin cases at later times. The aligned spins of the initial \ns[s] now result in higher angular velocities in the postmerger remnant core, increasing the $f_1$ and $f_2$ peaks of the postmerger \gw spectrum by $\sim O(\SI{50}{Hz})$.  Faster rotation also decreases the degree of differential rotation between the core and outer layers of the remnant, and thus the magnetic winding rate. The steepest gradients in the \pp rotation profile occur closer to the remnant surface than the \nsp case, leading to earlier activation of magnetic braking and a lower saturation value of $E_\mathrm{mag}$.

Amplification of the initial magnetic field leads to a rapid eruption from the inner accretion disk that launches a polar outflow in the \nsp and \pp cases. This eruption, marked by a sharp increase in the polar mass ejection rate and Poynting luminosity, coincides with the activation of the \parker in the polar funnel. Due to the reduced debris pollution mentioned above, the eruption occurs earlier in the \pp case and accelerates a faster, denser, and more magnetized outflow. Inferred terminal Lorentz factors remain low ($\lesssim2$), rendering the outflows inconsistent with \sgrb[s] at $\SI{60}{ms}$ postmerger, even in the \pp case where high initial magnetic fields, neutrino cooling of the disk, and a clean polar funnel create ideal conditions for jet launching. However, previous numerical and analytic studies of spinning magnetars show neutrino cooling to amplify jet energetics after much longer timescales of $1-\SI{10}{s}$ by further flattening the accretion disk and reducing neutrino-driven winds polluting the outflow, eventually achieving $\Gamma_\infty \gtrsim 100$ \cite{metzger_protomagnetar_2011, metzger_magnetar_2018, desai_relativistic_2026}. Future \grmhd + M1 simulations extending to longer postmerger times will be required to capture the evolution of $\Gamma_\infty$ over these timescales.

Due to the strong neutrino reprocessing captured by an M1 scheme, these magnetically driven outflows are uniquely proton rich, each containing $\SI{2.4e-3}{M_\odot}$ of material with $Y_e \geq 0.49$ and producing light r-process elements in the iron group and first peak. We infer this ejecta component to consistently appear in long-lived \bns merger remnants and provide a robust source of $^{56}\mathrm{Ni}$, whose emission after a half-life of $6$ days can lead to a flattening of the \kNe light curve decay and serve as a strong indication of neutrino- and magnetic-driven activity in the vicinity of the remnant. On the other hand, the dynamical ejecta remained neutron-rich in all cases, effectively reproducing observed solar system r-process abundances of heavy elements in the second and third peaks.

We note that such a strong link between proton-rich outflows and long-lived remnants relies on several assumptions and model limitations present in this study. The use of strong, initial dipolar magnetic fields with reflection symmetry can overestimate the amount of magnetic energy contained in small wavenumbers, and subsequently underestimate the time required to reorganize the turbulent field at merger into large scale coherent structures \cite{aguilera-miret_delayed_2024, gutierrez_magnetic_2025}. Additionally, improved microphysics via Monte Carlo neutrino transport \cite{foucart_robustness_2024}, inclusion of neutrino flavor oscillations \cite{qiu_neutrino_2025, qiu_impact_2025, lund_angle-dependent_2025}, or independent evolution of muon neutrino species \cite{ng_accurate_2025} (however, see \cite{gieg_consistent_2026} for a counterexample where muons have a limited impact) can all shift the late-time polar outflow towards lower $Y_e$. Future high resolution studies with more realistic microphysics coupled to detailed \kNe postprocessing are required to constrain the production of proton-rich outflows and clarify the impact of remnant lifetime on the \kNe signal.

\section{Acknowledgements}

The authors thank Leonardo Werneck, Zachariah Etienne, David Radice, Zhenyu Zhu, Luciano Combi, and Brian Metzger for valuable comments and discussions. A.W. gratefully acknowledges NASA support from FINESST grant 80NSSC25K0307. A.W., J.V.K., M.Chabanov, M.Campanelli and Y.Z. gratefully acknowledge support from NSF grant OAC-2411068 to UIUC and RIT, as well as NSF grants PHY-2409706 and PHY-2513442 to RIT, and NASA grant 80NSSC24K0100 to RIT. R.C. gratefully acknowledges support from the European Union under NextGenerationEU, via the PRIN 2022 Project “EMERGE'', Prot.~n.~2022KX2Z3B (CUP~C53D23001150006), and from INAF via the Theory Grant 2023 ``AfterJet'', Ob.Fu.~1.05.23.06.02 (CUP~C93C23006800005).

The authors acknowledge the Texas Advanced Computing Center (TACC) for providing computational resources on the Frontera supercomputer through allocations PHY-24027, PHY-20010 and AST-20021, as well as ACCESS for allocations PHY-260037 and PHY-240302. This research also used resources at the Rochester Institute of Technology (RIT), including the BlueSky, Green Prairies, and Lagoon clusters, which were acquired with support from NSF grants PHY-2018420, PHY-0722703, PHY-1229173, and PHY-1726215.

\appendix

\section{Analysis Methods}
\label{ax:analysis}

\subsection{Averaged Radial Profiles}
When the postmerger system approaches an axisymmetric state, it becomes informative to compute averaged radial profiles of remnant characteristics. Given a quantity $U$ defined on the computational domain, we compute its azimuthal and time-average as
\begin{equation}
    \langle U(R, t_\mathrm{avg})\rangle = \frac{1}{2\pi\Delta t} \int_0^{2\pi} \int_{t_\mathrm{avg} - \Delta t/2}^{t_\mathrm{avg} + \Delta t/2} U(R,\,\phi,\,t)\,dt\,d\phi
    \label{eq:phiavg}
\end{equation}
in the equatorial plane, where $R$ measures the cylindrical radius from the center of mass. We use $\Delta t = \SI{0.5}{ms}$ throughout the paper.

\subsection{Unbound Material}
We place a spherical detector at a radius of $R_D = \SI{740}{km}$ and interpolate \grmhd quantities onto its grid of $(n_\theta,\,n_\phi) = (55,\,96)$ points to track properties of unbound material. A fluid element interpolated to a detector point is flagged as unbound if it satisfies the the Bernoulli criterion
\begin{equation}
    -hu_t > h_\infty
    \label{eq:bernoulli}
\end{equation}
where $h$ is the specific enthalpy. This assumes efficient conversion from thermal to kinetic energy along the geodesic traversed by a fluid element, and may overestimate the amount of unbound material.  $h_\infty$ is taken to be the minimum enthalpy in the \eos table to further include the release of nuclear binding energy following the nucleosynthesis that occurs at late times \cite{fujibayashi_postmerger_2020}. In practice, $h_\infty \approx 1$ for the APRLDP \eos used here. The resulting asymptotic velocity is
\begin{equation}
    v_\infty = \sqrt{1 - \frac{1}{(-hu_t/h_\infty)^2}}
    \label{eq:vinf}
\end{equation}
The Bernoulli criterion is distinct from the geodesic criterion, which flags material as unbound if the asymptotic velocity is nonzero assuming $u_t$ is conserved, that is, if $-u_t > 1$. See, for example, \cite{bovard_r-process_2017, combi_grmhd_2023} and references within for discussions and comparisons between the two criteria.

We calculate the instantaneous mass ejection rate of unbound material as an integral over the detector surface $S$:

\begin{equation}
\begin{split}
    \dot{M}_\mathrm{ej} & = \int_S \mathcal{T}(\theta,\,\phi)\sqrt{\gamma}\Gamma\rho(\alpha v^r - \beta^r)dA \\
    & \equiv \int_S \dot{\mathcal{M}}_\mathrm{ej}(\theta,\,\phi)dA
\end{split}
\end{equation}
where $\gamma$ is the determinant of the 3-metric, $\Gamma$ is the fluid Lorentz factor, $\alpha$ and $\beta^i$ are the $3+1$ lapse and shift, $v^i$ is the Valencia 3-velocity, and $dA = R_D^2 \sin\theta\,d\theta d\phi$ is the surface element evaluated on the detector. $\mathcal{T}(\theta,\,\phi)$ is a flag for unbound material equal to one if the fluid element is unbound, and zero if bound. We also use the local mass ejection rate $\dot{\mathcal{M}}_\mathrm{ej}(\theta,\, \phi)$ as a weight for computing mass-averaged ejecta properties. For a quantity $U$, 
\begin{equation}
    \langle U\rangle = \frac{1}{T}\int_{0}^{T} \left( \frac{1}{\dot{M}_\mathrm{ej}}\int_S \dot{\mathcal{M}}_\mathrm{ej}(\theta,\, \phi) U(\theta,\, \phi) \,dA \right)~dt
    \label{eq:mass_avg}
\end{equation}
is the average taken over unbound material crossing the detector until time $T$.

We note that this detector is also used to record the outward \elmag Poynting and neutrino flux, as well as the spherical harmonic decomposition of $\psi_4$ used to calculate the \gw strain via fixed-frequency integration. Fluxes are then integrated over the surface to calculate the luminosity.

\subsection{Tracer Particles and Nucleosynthesis}
We embed 96000 passive Lagrangian tracer particles into the fluid at $t=0$ to record the evolution of $\rho,\,T,$ and $Y_e$ within unbound fluid elements. These data are later used as input to a nuclear reaction network to compute total nucleosynthetic yields. Initial particle locations are selected randomly from regions of the domain above a density threshold, placing them within the initial \ns[s]. The resulting particle distribution is uniform over the coordinate volume occupied by the \ns[s]. Each particle is then assigned a weight according to the fluid density at its initial position, following the method in \cite{curtis_magnetized_2024}. Particles are advected with the fluid after each simulation timestep with a forward Euler scheme:
\begin{equation}
    x^i_{t+\Delta t} = x^i_t + \Delta t (\alpha v^i - \beta^i)
    \label{eq:pt_evol}
\end{equation}
where $\Delta t$ is the timestep of the finest level. 

\begin{figure}
    \centering
    \includegraphics[width=\linewidth]{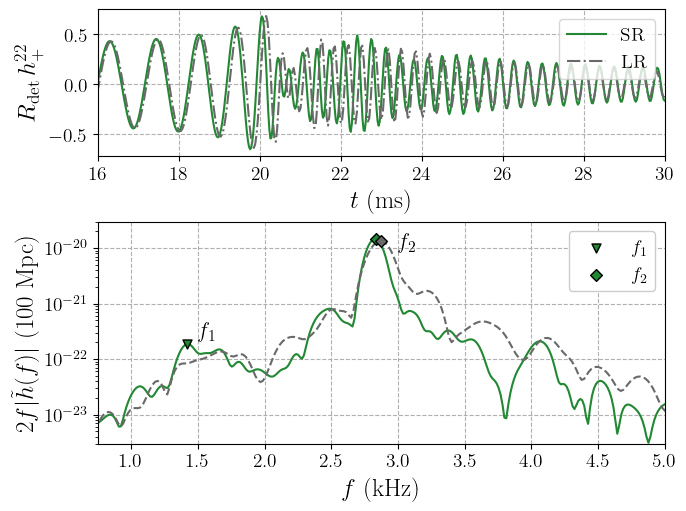}
    \caption{Differences in the $l=2,\,m=2$ \gw strain (top) and total \gw characteristic frequency strain (bottom) across resolution cases. Data from the SR simulation is shown in green solid curves, while the LR simulation is shown in gray dash-dotted curves. Note that the strain data has not been shifted in time, unlike Fig.~\ref{fig:gws}.}
    \label{fig:res_GWs}
\end{figure}

At the end of each simulation, all tracer particles remain on the grid. We filter for unbound tracers using the Bernoulli criterion (Eq.~(\ref{eq:bernoulli})) and find the iteration when the temperature at the tracer falls below $\SI{10}{GK}$ for the last time. We then pass the trajectory of $\rho(t),\,T(t),$ and $Y_e(t)$ after this time to the nuclear reaction network code \skynet \cite{lippuner_skynet_2017}. We extrapolate particle trajectories beyond the end of simulated data assuming homologous expansion. \skynet initializes a \nse state at the start of the trajectory and evolves the composition of 7836 isotopes, from free neutrons to $^{337}\mathrm{Cn}$, in \nse until cooling to $\SI{7}{GK}$, at which the full reaction network is used. We employ the same input nuclear physics as \cite{lippuner_r-process_2015}, using nuclear reaction rates and masses from the {\sc JINA REACLIB} database \cite{cyburt_jina_2010}.

\section{Effect of Resolution}
\label{ax:res}

\begin{figure}
    \centering
    \includegraphics[width=\linewidth]{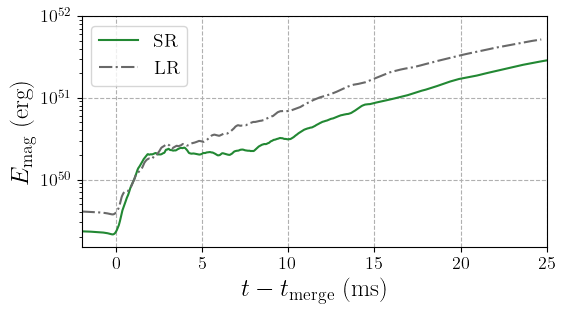}
    \caption{Time evolution of total magnetic energy for SR and LR simulations.}
    \label{fig:res_Emag}
\end{figure}

\begin{figure}
    \centering
    \includegraphics[width=\linewidth]{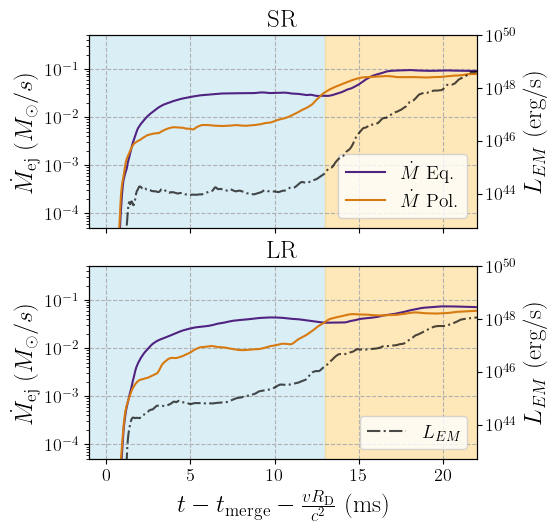}
    \caption{We plot the same quantities as Fig.~\ref{fig:Mdot_L} for \nsp SR (top) and LR (bottom) simulations.}
    \label{fig:res_Mdot}
\end{figure}

As described in Section \ref{ssec:grid}, here we compare nonspinning simulations at SR and LR. They differ in maximum resolution, where $\Delta x_{SR} = \SI{177}{m}$ and $\Delta x_{LR} = \SI{223}{m}$, and initial magnetic field strength, where $E_\mathrm{mag,\,SR}(t=0) = \SI{2.5e49}{erg}$ and $E_\mathrm{mag,\,LR}(t=0) = \SI{5e49}{erg}$. Figure~\ref{fig:res_GWs} shows the dependence of the \gw signal on resolution. In the top panel, we show the $l=2,\,m=2$ \gw strain with $t=0$ at initial data, indicating that the SR simulation reaches merger $\SI{0.13}{ms}$ faster than with LR. While this difference is small and indicates comparable inspiral dynamics across resolution, most resolution studies of \bns inspirals and associated \gw[s] show the opposite trend, where increased resolution decreases numerical dissipation of angular momentum, prolonging the inspiral \cite{radice_beyond_2013, kiuchi_toward_2025}. In the bottom panel, we show the characteristic frequency strain as was done in Fig.~\ref{fig:gw_ffts}. There is no obvious $f_1$ peak in the LR spectrum, emphasizing the uncertain nature of its physical meaning and appearance. The $f_2$ peak is $\SI{35}{Hz}$ higher in the LR simulation, which sets the approximate uncertainty in $f_2$ due to resolution.

\begin{figure}
    \centering
    \includegraphics[width=\linewidth]{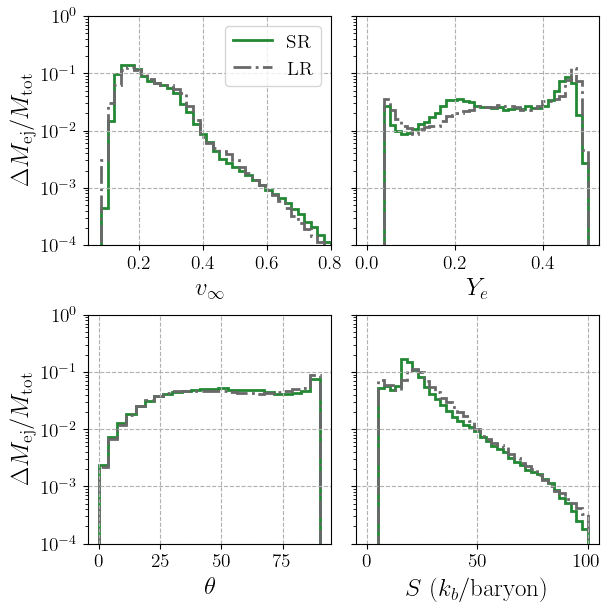}
    \caption{Distributions of mass ejected prior to $\SI{24}{ms}$ postmerger for SR and LR simulations.}
    \label{fig:res_hists}
\end{figure}

Despite starting from a higher initial $E_\mathrm{mag}$, the LR case encounters greater numerical dissipation in the turbulent \khi phase, after which the $E_\mathrm{mag}$ is comparable to the SR case, as shown at $\SI{4}{ms}$ postmerger in Fig.~\ref{fig:res_Emag}. Additionally, we show the resulting differences in mass ejection rate and \elmag luminosity across resolutions in Fig.~\ref{fig:res_Mdot}.
The LR case shows greater $L_\mathrm{EM}$ in the breakout phase by a factor of 2, following the initial $E_\mathrm{mag}$ being twice as high, while $\dot{M}_\mathrm{ej}$ remains comparable, suggesting that $L_\mathrm{EM}$ is sourced from magnetic field advected with dynamical ejecta prior to eruption. Both cases transition to the eruption phase at roughly the same time. The eruption in the SR case causes a more dramatic increase of $\dot{M}_\mathrm{ej}$ and $L_\mathrm{EM}$, reaching comparable, but higher, values compared to the LR case despite starting from a lower initial state. This implies that while the mechanism for buoyant magnetic eruption is consistent across resolution, the ability of this mechanism to initiate a distinct late-time steady state is less affected by the magnitude of $E_\mathrm{mag}$ contained within the remnant, and more sensitive to how turbulent processes in the disk are resolved. 

As the LR simulation ends $\SI{24}{ms}$ postmerger, before the steady state magnetar wind develops, the mass histograms in Fig.~\ref{fig:res_hists} are dominated by dynamical ejecta. By only considering material ejected in the SR simulation up to this point, we see similar distributions, suggesting that resolution does not add major uncertainties to our analysis of dynamical ejecta.

\section{Inspiral Dynamics and $\downarrow\downarrow$ Pre-Merger Treatment}
\label{ax:mm}

\begin{figure}
    \centering
    \includegraphics[width=\linewidth]{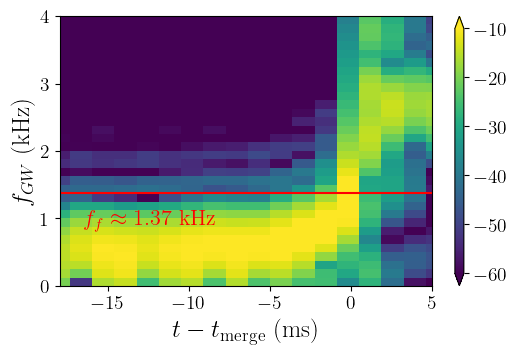}
    \caption{Spectrogram of the \mm \gw signal during the inspiral phase. The amplitude scale is arbitrary. We mark the shifted $f$-mode frequency estimated from Eq. (\ref{eq:omegaf}) with a red line. The initial time bins where the signal is dominated by junk radiation from initial data are not shown.}
    \label{fig:chirp}
\end{figure}

\begin{figure}
    \centering
    \includegraphics[width=\linewidth]{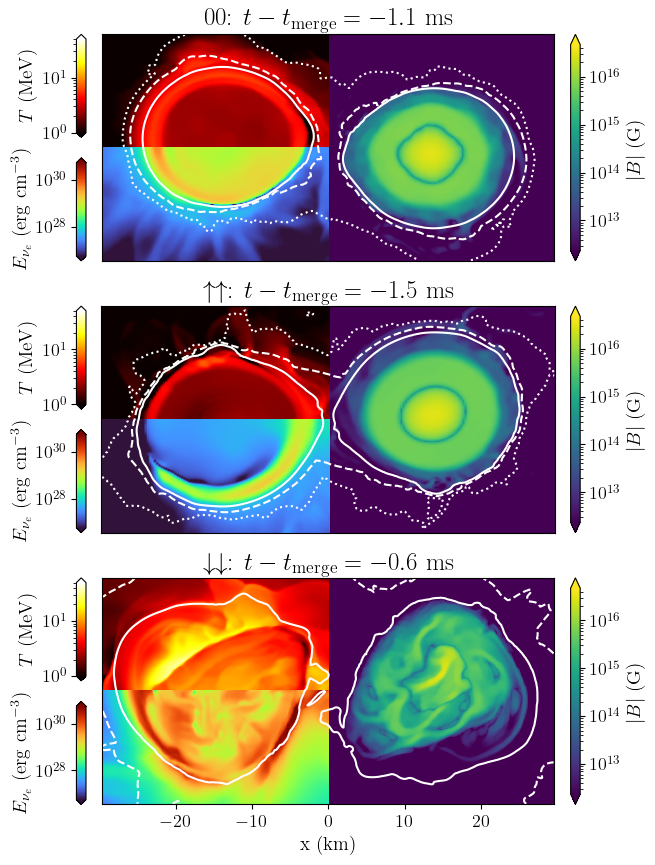}
    \caption{The late inspiral of each simulation, shown at a separation of $\sim\SI{28}{km}$. The top, middle, and bottom rows show the \nsp, \pp, and \mm simulations, respectively. The magnetic field magnitude $|B|$ is shown on the right, and the temperature $T$ and electron neutrino energy density $E_{\nu_e}$ are shown on the upper and lower left. Density contours mark values of $6.3 \times 10^{7},\, 6.3\times 10^{9},$ and $\SI{6.3e11}{g~cm}^{-3}$. Note that the lowest density contour is not visible in the \mm plot.}
    \label{fig:inspiral}
\end{figure}

We first estimate when the tidal resonance should occur in the \mm inspiral. Following \cite{kuan_tidal_2025}, we use a universal relation for the $f$-mode frequency shift due to \ns spin from \cite{kruger_fast_2020}:
\begin{equation}
    \frac{f_f}{f_0} = 1 - 0.235 \left(\frac{\Omega}{\Omega_K}\right)-0.491\left(\frac{\Omega}{\Omega_K}\right)^2
    \label{eq:omegaf}
\end{equation}
where $\Omega$ is the angular velocity of the \ns, $\Omega_K$ is the Kepler (mass shedding) velocity, and $f_0$ is the $f$-mode frequency of a nonspinning star. We use additional relations from \cite{doneva_gravitational_2013} to estimate the latter two parameters:
\begin{equation}
    \frac{1}{2\pi}\Omega_K~[\mathrm{kHz}] = 1.716 \sqrt{\frac{M/\SI{1.4}{M_\odot}}{(R/\SI{10}{km})^3}} - 0.189
    \label{eq:omegaK}
\end{equation}
\begin{equation}
    f_0~[\mathrm{kHz}] = 1.151 \sqrt{\frac{M/\SI{1.4}{M_\odot}}{(R/\SI{10}{km})^3}} +1.562
    \label{eq:omega0}
\end{equation}
where $M$ is the \ns mass and $R$ is the radius of a nonspinning \ns of mass $M$. We find $\Omega_K \approx \SI{1.02}{kHz}$ and $f_0 \approx \SI{2.37}{kHz}$ using our \mm \bns parameters. Equation \ref{eq:omegaf} then gives us $f_f \approx \SI{1.37}{kHz}$. The spectrogram in Fig.~\ref{fig:chirp} shows the change in the \gw frequency spectrum from the inspiral to merger.  This allows for a brief window in the final $\sim \SI{1}{ms}$ before merger where resonance can be achieved.

The consequences of these tidal interactions are clearer in the structure of the individual \ns[s] prior to merger. To clarify the inspiral dynamics and motivate the additional procedures used to stabilize the \mm simulation referenced in Section \ref{ssec:id}, Fig.~\ref{fig:inspiral} shows the three binaries when separated by $\sim \SI{28}{km}$. The snapshot of the \mm simulation in the bottom row occurs \SI{0.2}{ms} \textit{after} magnetic fields are superimposed on the \mm \ns[s], but \SI{0.4}{ms} \textit{before} neutrino fields are reset to equilibrium. While the magnetic fields set in the \nsp and \pp simulations at $t=0$ retain their initial dipolar structures, internal motions within the \mm \ns[s] immediately rearrange the prescribed field. Density contours also show the geometric distortions of the \mm \ns[s], in contrast to the more ellipsoidal \nsp and \pp \ns[s] (see \cite{steinhoff_spin_2021} for similar deformations in a BH-NS binary with antialigned spins).

\begin{figure}
    \centering
    \includegraphics[width=\linewidth]{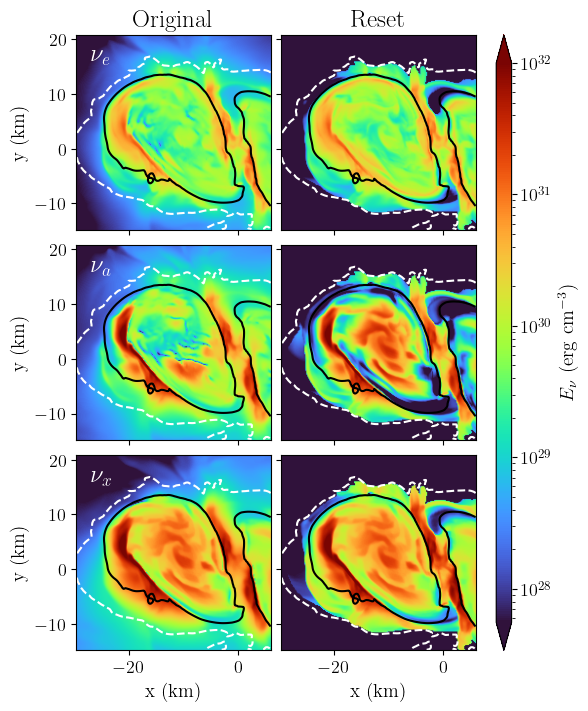}
    \caption{Neutrino energy densities in the \mm simulation $\SI{0.2}{ms}$ before merger, before (left) and after (right) they are reset to $\beta$-equilibrium. Electron neutrino, electron antineutrino, and heavy neutrino energies are shown in the top, middle, and bottom rows, respectively. Density contours mark values of $6.3\times 10^{10}$ and $\SI{6.3e12}{g~cm}^{-3}$.}
    \label{fig:nureset}
\end{figure}

The \mm case is also hotter, with more energetic neutrino fields in Fig.~\ref{fig:inspiral}, compared to the other simulations. When this neutrino field is subjected to the subsequent violent merger, the M1 algorithm will fail to find $\beta$-equilibrium where it is enforced for reaction rate calculations in optically thick regions (see Sec. 3.2.2 in \cite{radice_new_2022}). To avoid this issue while preserving the hydrodynamic state from Fig.~\ref{fig:inspiral}, we choose to reset neutrino radiation fields to $\beta$-equilibrium immediately before merger. The neutrino energy densities before and after the reset are shown in Fig.~\ref{fig:nureset}. While changes to the electron and heavy neutrino fields are minimal within the bulk of the \ns[s], electron antineutrino energies are reduced near the \ns surface and increased near the core. These changes are small compared to physical neutrino emission at merger, where neutrino energies reach values of $\sim\SI{3e35}{erg~cm}^{-3}$ in the hot shear layer and $\sim\SI{4e34}{erg~cm}^{-3}$ within the remnant.

\clearpage
\bibliographystyle{apsrev4-2}
\bibliography{references}
\end{document}